\documentclass[conference]{IEEEtran}
\IEEEoverridecommandlockouts
\usepackage{cite}
\usepackage{amsmath,amssymb,amsfonts}
\usepackage{algorithmic}
\usepackage{graphicx}
\usepackage{textcomp}
\usepackage{xcolor}
\usepackage[a4paper, total={184mm,239mm}]{geometry}
\def\BibTeX{{\rm B\kern-.05em{\sc i\kern-.025em b}\kern-.08em
    T\kern-.1667em\lower.7ex\hbox{E}\kern-.125emX}}

\usepackage[breaklinks,hidelinks]{hyperref}
\usepackage{circledsteps}

\usepackage{caption}
\usepackage{subcaption}
\usepackage{multirow}
\usepackage{balance}

\usepackage{url}

\usepackage{listings}

\lstdefinestyle{rwth-style}{
    basicstyle=\ttfamily\footnotesize,
    breakatwhitespace=true,
    breaklines=true,
    captionpos=b,
    commentstyle=\color{grun},
    escapechar=\%,
    floatplacement=htb,
    frame=tb,
    framexleftmargin=15pt,
    keepspaces=true,
    keywordstyle=\color{rwth},
    language=C++,
    numbers=left,
    numbersep=8pt,
    numberstyle=\tiny\color{black-75},
    numberblanklines=true,
    showspaces=false,
    showstringspaces=false,
    showtabs=false,
    stringstyle=\color{bordeaux},
    tabsize=2,
    xleftmargin=2em,
    belowskip=-0.8\baselineskip,
}
\usepackage{enumitem}
\setlist[itemize]{noitemsep, topsep=0pt}
\setlist[enumerate]{noitemsep, topsep=0pt}

\usepackage{tikz}
\usetikzlibrary{
    positioning,
    shapes,
    arrows,
    arrows.meta,
    matrix,
    fit,
    patterns,
    patterns.meta,
    calc
}

\usepackage{pgfplots}
\pgfplotsset{compat=1.18}
\usepackage{pgfplotstable}

\pgfplotsset{
    y axis style/.style={
        yticklabel style=#1,
        ylabel style=#1,
        y axis line style=#1,
        ytick style=#1
    },
    General/.style={
        font=\footnotesize,
        width=\linewidth,
        height=3.2cm,
        xtick pos=left,
        xtick align=outside,
        ytick pos=left,
        ytick align=outside,
        ymajorgrids=true,
        grid style=dashed,
        legend cell align={left},
        style=thin,
    },
    BarConfig/.style={
        General,
        ymin=0,
        enlarge x limits=0.1,
        xtick=data,
        bar width=8pt,
    },
    ScatterConfig/.style={
      General,
      only marks,
      xtick=data,
      xticklabel style={rotate=90, anchor=east},
    }
}

\usepackage[pages=some]{background}

\usepackage[printonlyused,nolist]{acronym}
\usepackage[capitalise,nameinlink]{cleveref}
\usepackage{siunitx}
\usepackage{orcidlink}

\makeatletter
\renewcommand*\AC@acs[1]{%
    \expandafter\AC@get\csname fn@#1\endcsname\@firstoftwo{#1}}
\makeatother

\def\BibTeX{{\rm B\kern-.05em{\sc i\kern-.025em b}\kern-.08em
    T\kern-.1667em\lower.7ex\hbox{E}\kern-.125emX}}
\begin{document}
\bstctlcite{IEEEexample:BSTcontrol}

\begin{acronym}[AAPCS64]
    \acro{aapcs64}[AAPCS64]{Procedure Call Standard for the \acl{aarch64}}
    \acro{aarch64}[AArch64]{\acs{arm} 64-bit Architecture}
    \acro{ai}[AI]{Artificial Intelligence}
    \acro{aiba}[AIBA]{An Automated Intra-cycle Behavioral Analysis for SystemC-based design exploration}
    \acro{algol60}[ALGOL~60]{Algorithmic Language 1960}
    \acro{amd}[AMD]{Advanced Micro Devices}
    \acro{aoa}[AoA]{ARM-on-ARM}
    \acro{api}[API]{Application Programming Interface}
    \acro{arm}[ARM]{Advanced \acs{risc} Machines}
    \acro{ast}[AST]{Abstract Syntax Tree}
    \acro{avp64}[AVP64]{\acs{arm}v8 \acl{vp}}
    \acro{bar}[BAR]{Base Address Register}
    \acro{bb}[BB]{Basic Block}
    \acro{bl}[BL]{Branch-With-Link}
    \acro{bp}[BP]{Breakpoint}
    \acro{can}[CAN]{Controller Area Network}
    \acro{ci}[CI]{Continous Integration}
    \acro{cicd}[CI/CD]{Continuous Integration/Continuous Delivery}
    \acro{clint}[CLINT]{Core-Local Interrupt Controller}
    \acro{cpu}[CPU]{Central Processing Unit}
    \acro{crc}[CRC]{Cyclic Redundancy Check}
    \acro{csv}[CSV]{Character-Separated Values}
    \acro{db}[DB]{Database}
    \acro{dbt}[DBT]{Dynamic Binary Translation}
    \acro{dbms}[DBMS]{\Acl{db} Management System}
    \acro{ddr}[DDR]{Double Data Rate}
    \acro{des}[DES]{Discrete Event Simulation}
    \acro{dla}[DLA]{Deep Learning Accelerator}
    \acro{dma}[DMA]{Direct Memory Access}
    \acro{dmi}[DMI]{Direct Memory Interface}
    \acro{ds}[DS]{Developer Studio}
    \acro{dwarf}[DWARF]{Debugging With Arbitrary Record Formats}
    \acro{eda}[EDA]{Electronic Design Automation}
    \acro{eembc}[EEMBC]{Embedded Microprocessor Benchmark Consortium}
    \acro{etrace}[etrace]{Execution Trace}
    \acro{el}[EL]{Exception Level}
    \acro{elf}[ELF]{Executable and Linkable Format}
    \acro{elog}[elog]{Execution Log}
    \acro{esa}[ESA]{European Space Agency}
    \acro{esl}[ESL]{Electronic System Level}
    \acro{fd}[fd]{file descriptor}
    \acro{fig}[FIG]{Fault Injection in glibc}
    \acro{fp}[FP]{Frame Pointer}
    \acro{fpga}[FPGA]{Field Programmable Gate Array}
    \acro{fss}[FSS]{Full-System Simulator}
    \acro{fvp}[FVP]{Fixed Virtual Platform}
    \acro{gcc}[GCC]{GNU Compiler Collection}
    \acro{gdb}[GDB]{GNU Debugger}
    \acro{gic}[GIC]{Generic Interrupt Controller}
    \acro{got}[GOT]{Global Offset Table}
    \acro{gpio}[GPIO]{General Purpose Input/Output}
    \acro{gpu}[GPU]{Graphics Processing Unit}
    \acro{gui}[GUI]{Graphical User Interface}
    \acro{hart}[hart]{Hardware Thread}
    \acro{html}[HTML]{HyperText Markup Language}
    \acro{hw}[HW]{Hardware}
    \acro{ibm}[IBM]{International Business Machines}
    \acro{id}[ID]{identifier}
    \acro{ieee}[IEEE]{Institute of Electrical and Electronics Engineers}
    \acro{io}[I/O]{Input/Output}
    \acro{iommu}[IOMMU]{Input-Output Memory Management Unit}
    \acro{iot}[IoT]{Intenet of Things}
    \acro{iova}[IOVA]{IO Virtual Address}
    \acro{ipa}[IPA]{Intermediate Physical Addresses}
    \acro{ir}[IR]{Intermediate Representation}
    \acro{irq}[IRQ]{Interrupt Request}
    \acro{isa}[ISA]{Instruction-Set Architecture}
    \acro{isa-bus}[ISA]{Industry Standard Architecture}
    \acro{iss}[ISS]{Instruction-Set Simulator}
    \acro{jit}[JIT]{Just-In-Time}
    \acro{kvm}[KVM]{Kernel Virtual Machine}
    \acro{lr}[LR]{Link Register}
    \acro{lt}[LT]{Loosely-Timed}
    \acro{mips}[MIPS]{Million Instructions Per Second}
    \acro{miso}[MISO]{Master Input Slave Output}
    \acro{mmio}[MMIO]{Memory-Mapped Input/Output}
    \acro{mmu}[MMU]{Memory Management Unit}
    \acro{mnist}[MNIST]{Modified National Institute of Standards and Technology}
    \acro{mosi}[MOSI]{Master Output Slave Input}
    \acro{msi}[MSI]{Message Signalled Interrupt}
    \acro{nas}[NAS]{\acs{nasa} Advanced Supercomputing}
    \acro{nasa}[NASA]{National Space Agency}
    \acro{nistt}[NISTT]{A Non-Intrusive SystemC-TLM 2.0 Tracing Tool}
    \acro{nn}[NN]{Neural Network}
    \acro{nop}[NOP]{No Operation}
    \acro{npb}[NPB]{\acs{nas} Parallel Benchmarks}
    \acro{nvdla}[NVDLA]{NVIDIA \acl*{dla}}
    \acro{openmp}[OpenMP]{Open Multi-Processing}
    \acro{os}[OS]{Operating System}
    \acro{pa}[PA]{Physical Address}
    \acrodefplural{pa}[PAs]{Physical Addresses}
    \acro{pc}[PC]{Program Counter}
    \acro{pccts}[PCCTS]{Purdue Compiler Construction Tool Set}
    \acro{pci}[PCI]{Peripheral Component Interconnect}
    \acro{pcie}[PCIe]{PCI Express}
    \acro{pid}[PID]{process ID}
    \acro{plic}[PLIC]{Platform-Level Interrupt Controller}
    \acro{plt}[PLT]{Procedure Linkage Table}
    \acro{pmu}[PMU]{Performance Monitoring Unit}
    \acro{pthread}[p\-thread]{\acs{posix} Thread}
    \acro{posix}[POSIX]{Portable Operating System Interface}
    \acro{qemu}[QEMU]{Quick Emulator}
    \acro{ram}[RAM]{Random-Access Memory}
    \acro{risc}[RISC]{Reduced Instruction Set Computer}
    \acro{rtc}[RTC]{Real-Time Clock}
    \acro{rtf}[RTF]{Real-Time Factor}
    \acro{rtl}[RTL]{Register-Transfer Level}
    \acro{rtos}[RTOS]{real-time operating system}
    \acro{sata}[SATA]{Serial AT Attachment}
    \acro{sd}[SD]{Secure Digital}
    \acro{sdhci}[SDHCI]{\acs{sd} Host Controller Interface}
    \acro{simd}[SIMD]{Single Instruction, Multiple Data}
    \acro{smmu}[SMMU]{System Memory Management Unit}
    \acro{soc}[SoC]{System-on-a-Chip}
    \acrodefplural{soc}[SoCs]{Systems-on-Chips}
    \acro{sp}[SP]{Stack Pointer}
    \acro{spi}[SPI]{Serial Peripheral Interface}
    \acro{sql}[SQL]{Structured Query Language}
    \acro{ssd}[SSD]{Solid State Drive}
    \acro{sw}[SW]{Software}
    \acro{tb}[TB]{Translation Block}
    \acro{tcg}[TCG]{Tiny Code Generator}
    \acro{tcp}[TCP]{Transmission Control Protocol}
    \acro{td}[TD]{Temporal Decoupling}
    \acro{tfl}[TFLite]{TensorFlow Lite}
    \acro{tlm}[TLM]{Transaction-Level Modeling}
    \acro{tpu}[TPU]{Tensor Processing Unit}
    \acro{uart}[UART]{Universal Asynchronous Receiver/Transmitter}
    \acro{unix}[UNIX]{Uniplexed Information and Computing Service}
    \acro{va}[VA]{Virtual Address}
    \acrodefplural{va}[VAs]{Virtual Addresses}
    \acro{vcd}[VCD]{Value Change Dump}
    \acro{vcml}[VCML]{Virtual Components Modeling Library}
    \acro{vcpu}[vCPU]{virtual CPU}
    \acro{vfio}[VFIO]{Virtual Function I/O}
    \acro{vhe}[VHE]{Virtualization Host Extensions}
    \acro{vpci}[vPCI]{virtual PCI}
    \acro{viper}[VIPER]{Virtual Platform Explorer}
    \acro{vm}[VM]{Virtual Machine}
    \acro{vp}[VP]{Virtual Platform}
    \acro{vt-d}[VT-d]{Virtualization Technology for Directed I/O}
    \acro{wfi}[WFI]{Wait For Interrupt}
\end{acronym}



\definecolor{rwth}   {RGB}{  0  84 159}
\definecolor{rwth-75}{RGB}{ 64 127 183}
\definecolor{rwth-50}{RGB}{142 186 229}
\definecolor{rwth-25}{RGB}{199 221 242}
\definecolor{rwth-10}{RGB}{232 241 250}

\definecolor{black}   {RGB}{  0   0   0}
\definecolor{black-75}{RGB}{100 101 103}
\definecolor{black-50}{RGB}{156 158 159}
\definecolor{black-25}{RGB}{207 209 210}
\definecolor{black-10}{RGB}{236 237 237}

\definecolor{magenta}   {RGB}{227   0 102}
\definecolor{magenta-75}{RGB}{233  96 136}
\definecolor{magenta-50}{RGB}{241 158 177}
\definecolor{magenta-25}{RGB}{249 210 218}
\definecolor{magenta-10}{RGB}{253 238 240}

\definecolor{yellow}   {RGB}{255 237   0}
\definecolor{yellow-75}{RGB}{255 240  85}
\definecolor{yellow-50}{RGB}{255 245 155}
\definecolor{yellow-25}{RGB}{255 250 209}
\definecolor{yellow-10}{RGB}{255 253 238}

\definecolor{petrol}   {RGB}{  0  97 101}
\definecolor{petrol-75}{RGB}{ 45 127 131}
\definecolor{petrol-50}{RGB}{125 164 167}
\definecolor{petrol-25}{RGB}{191 208 209}
\definecolor{petrol-10}{RGB}{230 236 236}

\definecolor{turkis}   {RGB}{  0 152 161}
\definecolor{turkis-75}{RGB}{  0 177 183}
\definecolor{turkis-50}{RGB}{137 204 207}
\definecolor{turkis-25}{RGB}{202 231 231}
\definecolor{turkis-10}{RGB}{235 246 246}

\definecolor{grun}   {RGB}{ 87 171  39}
\definecolor{grun-75}{RGB}{141 192  96}
\definecolor{grun-50}{RGB}{184 214 152}
\definecolor{grun-25}{RGB}{221 235 206}
\definecolor{grun-10}{RGB}{242 247 236}

\definecolor{maigrun}   {RGB}{189 205   0}
\definecolor{maigrun-75}{RGB}{208 217  92}
\definecolor{maigrun-50}{RGB}{224 230 154}
\definecolor{maigrun-25}{RGB}{240 243 208}
\definecolor{maigrun-10}{RGB}{249 250 237}

\definecolor{orange}   {RGB}{246 168   0}
\definecolor{orange-75}{RGB}{250 190  80}
\definecolor{orange-50}{RGB}{253 212 143}
\definecolor{orange-25}{RGB}{254 234 201}
\definecolor{orange-10}{RGB}{255 247 234}

\definecolor{rot}   {RGB}{204   7  30}
\definecolor{rot-75}{RGB}{216  92  65}
\definecolor{rot-50}{RGB}{230 150 121}
\definecolor{rot-25}{RGB}{243 205 187}
\definecolor{rot-10}{RGB}{250 235 227}

\definecolor{bordeaux}   {RGB}{161  16  53}
\definecolor{bordeaux-75}{RGB}{182  82  86}
\definecolor{bordeaux-50}{RGB}{205 139 135}
\definecolor{bordeaux-25}{RGB}{229 197 192}
\definecolor{bordeaux-10}{RGB}{245 232 229}

\definecolor{lila}   {RGB}{122 111 172}
\definecolor{lila-75}{RGB}{155 145 193}
\definecolor{lila-50}{RGB}{188 181 215}
\definecolor{lila-25}{RGB}{222 218 235}
\definecolor{lila-10}{RGB}{242 240 247}

\definecolor{violett}   {RGB}{ 97  33  88}
\definecolor{violett-75}{RGB}{131  78 117}
\definecolor{violett-50}{RGB}{168 133 158}
\definecolor{violett-25}{RGB}{210 192 205}
\definecolor{violett-10}{RGB}{237 229 234}

\newcommand{\uniquepgftag}{tag}
\newcommand{\uniquepgflabel}[1]{\label{\uniquepgftag#1}}
\newcommand{\uniquepgfref}[1]{\ref*{\uniquepgftag#1}}
\newcommand*\circled[1]{\tikz[baseline=(char.base)]{
    \node[shape=circle,draw,fill=white,inner sep=1pt] (char) {#1};}}

\tikzset{/csteps/fill color=white, /csteps/outer color=bordeaux-75, /csteps/inner color=black}

\title{High-Performance ARM-on-ARM Virtualization for Multicore SystemC-TLM-Based Virtual Platforms}

\author{
    \IEEEauthorblockN{
        Nils Bosbach\IEEEauthorrefmark{1}\orcidlink{0000-0002-2284-949X},
        Rebecca Pelke\IEEEauthorrefmark{1}\orcidlink{0000-0001-5156-7072},
        Niko Zurstraßen\IEEEauthorrefmark{1}\orcidlink{0000-0003-3434-2271},
        Jan Henrik Weinstock\IEEEauthorrefmark{2}\orcidlink{0009-0008-0902-7652},\\
        Lukas Jünger\IEEEauthorrefmark{2}\orcidlink{0000-0001-9149-1690},
        Rainer Leupers\IEEEauthorrefmark{1}\orcidlink{0000-0002-6735-3033}
    }
    \IEEEauthorblockA{
        \IEEEauthorrefmark{1}\textit{RWTH Aachen University}, Aachen, Germany\\
        \IEEEauthorrefmark{2}\textit{MachineWare GmbH}, Aachen, Germany\\
        \IEEEauthorrefmark{1}\{bosbach, pelke, zurstrassen, leupers\}@ice.rwth-aachen.de \quad
        \IEEEauthorrefmark{2}\{jan, lukas\}@mwa.re
    }
}

\newcommand\copyrighttext{%
  \footnotesize \textcopyright \the\year{} IEEE. Personal use of this material is permitted. Permission from IEEE must be obtained for all other uses, including reprinting/republishing this material for advertising or promotional purposes, collecting new collected works for resale or redistribution to servers or lists, or reuse of any copyrighted component of this work in other works.}

\newcommand\copyrightnotice{%
    \backgroundsetup{opacity=1, scale=1, angle=0, contents={
            \color{black}%
            \begin{tikzpicture}[remember picture,overlay]%
                \node[anchor=south,yshift=10pt] at (current page.south) {\fbox{\parbox{\dimexpr0.75\textwidth-\fboxsep-\fboxrule\relax}{\copyrighttext}}};
                \node[anchor=north,yshift=-10pt,text=gray] at (current page.north) {\shortstack[c]{\large PREPRINT - accepted by the \textit{Design, Automation and Test in Europe Conference 2025 (DATE '25)}\\DOI:\href{https://doi.org/10.23919/DATE64628.2025.10993216}{10.23919/DATE64628.2025.10993216}}};
            \end{tikzpicture}%
        }%
    }%
    \BgThispage%
}

\maketitle
\copyrightnotice

\crefname{figure}{Figure}{Figures}

\acused{cpu}
\acused{id}
\acused{tlm}
\begin{abstract}
    The increasing complexity of hardware and software requires advanced development and test methodologies for modern systems on chips.
    This paper presents a novel approach to \acl{aoa} virtualization within SystemC-based simulators using Linux's KVM to achieve high-performance simulation.

    By running target software natively on ARM-based hosts with hardware-based virtualization extensions, our method eliminates the need for instruction-set simulators, which significantly improves performance.
    We present a multicore SystemC-TLM-based CPU model that can be used as a drop-in replacement for an instruction-set simulator.
    It places no special requirements on the host system, making it compatible with various environments.

    Benchmark results show that our \acl{aoa}-based virtual platform achieves up to \SI{10}{x} speedup over traditional instruction-set-simulator-based models on compute-intensive workloads.
    Depending on the benchmark, speedups increase to more than \SI{100}{x}.
\end{abstract}
\acresetall


\acused{cpu}
\acused{id}
\acused{qemu}

\section{Introduction}
\label{sec:introduction}

Driven by Moore's Law, the complexity of \ac{hw} and \ac{sw} has steadily increased over the past decades.
This trend requires continuous evolution in the development and testing processes for modern \acp{soc}.
A technology that significantly supports companies in their \ac{sw} development is virtual prototyping.

Virtual prototyping involves creating a simulator of the entire system, referred to as a \ac{vp} or \ac{fss} that behaves like the \ac{soc} under design.
This \ac{vp} can serve as a target platform for \ac{sw} development while the \ac{hw} is still being designed.
One example is the pre-silicon development of device drivers using a virtual model of the device within a \ac{vp}.
\acp{vp} offer several other advantages including unlimited scalability, deep introspection, insightful tracing facilities, and seamless integration into automated \ac{cicd} workflows.

One critical aspect of a \ac{vp}'s usability is its performance.
High performance is essential for real-time debugging and automated testing.
Typically, one of the most compute-intensive models within a \ac{vp} is the \ac{cpu} model executing the target \ac{sw}.
When the host\footnote{host: architecture that runs the simulation}-machine architecture differs from that of the target\footnote{target: architecture to be simulated} system, an \ac{iss} within the \ac{cpu} model translates instructions from the target \ac{isa} to the host \ac{isa}.

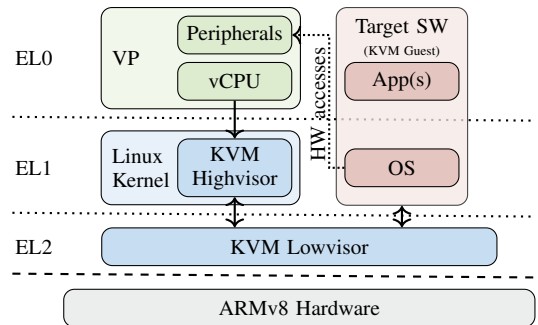
\begin{figure}[!t]
    \centering
    \tikzstyle{nodetype} = [rectangle, rounded corners, minimum width=1.2cm, minimum height=.5cm, text centered, text width=1.3cm, draw=black, fill=rwth-25, font={\footnotesize}, inner sep=0.1cm]
\tikzstyle{lbl}      = [font={\footnotesize}, inner sep=1pt]

\pgfdeclarelayer{background}
\pgfdeclarelayer{middle}
\pgfsetlayers{background,middle,main}

\begin{tikzpicture}
    \node[nodetype, fill=black-10, text width=6.0cm] (hw) {ARMv8 Hardware};

    \node[nodetype, text width=5.0cm, anchor=south] at ([yshift=+0.3cm]hw.north) (kvm-lv) {\acs{kvm} Lowvisor};

    \node[nodetype, anchor=south west] at ([xshift=1.0cm,yshift=+0.4cm]kvm-lv.north west) (kvm-hv) {\acs{kvm} Highvisor};
    \node[lbl, text width=0.9cm, anchor=west] at ([xshift=0.1cm]kvm-lv.west|-kvm-hv.center) (linux) {Linux Kernel};
    \begin{pgfonlayer}{background}
        \node[nodetype, fit={(kvm-hv) (linux)}, fill=rwth-10] (linux-box) {};
    \end{pgfonlayer}

    \node[nodetype, anchor=south, fill=grun-25] at ([yshift=0.5cm]kvm-hv.north) (vcpu) {\acs{vcpu}};
    \node[nodetype, anchor=south, fill=grun-25] at ([yshift=0.1cm]vcpu.north) (vhw) {Peripherals};
    \node[lbl, inner sep=0pt, fit={(vcpu) (vhw)}] (vp_elem) {};
    \node[lbl, text width=0.9cm, anchor=west] at (linux.west|-vp_elem) (vp) {\acs{vp}};
    \begin{pgfonlayer}{background}
        \node[nodetype, fit={(vp_elem) (vp)}, fill=grun-10] (vp-box) {};
    \end{pgfonlayer}

    \node[nodetype, fill=bordeaux-25, anchor=east] at ([xshift=-0.5cm]linux-box.center-|kvm-lv.east) (target-os) {\acs{os}};

    \node[nodetype, fill=bordeaux-25, anchor=center] at (vcpu.center-|target-os.center) (target-app) {App(s)};

    \phantom{\node[lbl, inner sep=0pt, fit={(target-os) (target-app) (target-app|-vhw.north)}] (target-phantom) {};}
    \node[lbl, anchor=north] at (target-phantom|-vhw.north) (target-sw) {\shortstack{Target \ac{sw}\\\tiny(\acs{kvm} Guest)}};
    \begin{pgfonlayer}{middle}
        \node[nodetype, fit={(target-sw) (target-phantom) ([yshift=0.1cm]target-os.south|-linux-box.south)}, fill=bordeaux-10, nearly opaque] (target-box) {};
    \end{pgfonlayer}

    \begin{pgfonlayer}{background}
        \node[lbl, anchor=east] at ([xshift=-0.1cm]vp-box.center-|hw.west) (el0) {\acs{el}0};
        \node[lbl, anchor=west] at (linux-box.center-|el0.west) (el1) {\acs{el}1};
        \node[lbl, anchor=west] at (kvm-lv.center-|el0.west) (el2) {\acs{el}2};
        \draw[dotted, thick] ([yshift=-0.1cm]el0.west|-vp-box.south) -- ([yshift=-0.15cm]vp-box.south-|hw.east);
        \draw[dotted, thick] ([yshift=-0.1cm]el0.west|-linux-box.south) -- ([yshift=-0.15cm]linux-box.south-|hw.east);
        \draw[dashed, thick] ([yshift=-0.1cm]el0.west|-kvm-lv.south) -- ([yshift=-0.15cm]kvm-lv.south-|hw.east);
    \end{pgfonlayer}

    \draw[->, thick] (vcpu) -- (kvm-hv);
    \draw[<->, thick] (kvm-hv.south) -- (kvm-hv.center|-kvm-lv.north);
    \draw[<->, thick] (target-box.south) -- (kvm-lv.north-|target-box.center);
    \draw[->, thick, densely dotted] (target-os.west) -- ++(-0.2cm, 0cm) |- node[lbl, above, pos=0.01, rotate=90, anchor=south west] {HW accesses} (vhw.east);

\end{tikzpicture}
    \caption{\acs{vcpu} integration using \acs{kvm}.}
    \label{fig:kvm}
\end{figure}

However, if both architectures match and \ac{vhe}, such as ARMv8's \ac{vhe}~\cite{arm_ltd_arm_2024}, are available on the host, the target \ac{sw} can be executed natively without the need for an \ac{iss}.
\Cref{fig:kvm} illustrates how Linux's hypervisor \ac{kvm} can be used by the \ac{vcpu} to run the target \ac{sw} natively on the host.
Accesses to memory-mapped peripherals are trapped by \ac{kvm} and forwarded to the \ac{cpu} model running in the \ac{vp} process on the host.
This ensures that the target \ac{sw} accesses the virtual \ac{hw} instead of the physical host \ac{hw}.

The industry-standard framework for building \acp{vp} is SystemC~\cite{systemc-2023}.
SystemC is a C++ library providing standardized interfaces that are crucial for compatibility across different simulations and tools.
\ac{tlm} further extends the standard by providing abstract interfaces to model communication between different modules.

Building on previous work that integrated \ac{kvm} into a SystemC-based \ac{cpu} model~\cite{junger_arm--arm_2020}, we add multicore support and eliminate limitations this approach has.
With Apple's transition from Intel-based x86 to ARM-based Apple Silicon processors in 2020~\cite{iyengar_apple_2020}, ARM-based \ac{hw} that can be used as a platform for \ac{aoa}-based \acp{vp} has become widely available.

In this paper, we present:

\begin{itemize}
    \item A multicore, SystemC-\ac{tlm}-based \ac{cpu} model that uses \ac{kvm} and runs the simulated cores in parallel on the host.
        It serves as a drop-in replacement for an \ac{iss} without requiring any other adjustments to the \ac{vp}.
    \item An implementation independent of performance counters.
    \item An alternate approach to custom kernel patches enhancing simulation performance.
    \item Benchmark results demonstrating our \ac{aoa}-based \ac{vp}'s effectiveness compared to traditional \acp{iss}.
\end{itemize}

\section{Background \& Related Work}
\label{sec:background}

The field of virtual prototyping for embedded \ac{sw} development has seen significant advancements in recent years, particularly in the context of ARM-based virtualization and SystemC-based simulations.
This section provides an overview of existing literature and research in this domain, highlighting key contributions but also limitations of previous work.

\subsection{Virtual Prototyping for Software Development}
Nowadays, the creation of \acp{vp} during the design process of an \ac{soc} has become essential.
\acp{vp} model the behavior of the full system allowing unmodified target \ac{sw} to be executed.
This facilitates debugging and tracing~\cite{bosbach_nistt_2022} of the target \ac{sw}.

SystemC has emerged as the industry-standard framework for building \acp{vp} due to its standardized interfaces and compatibility across different simulations and tools~\cite{systemc-2023}.
It provides basic module classes, a scheduler to simulate parallelism, and a concept of time.
\ac{tlm} extends SystemC by providing abstract interfaces to model communication between modules.

To further extend SystemC's features, modeling libraries are available that add frequently needed parts, components, and convenience functions.
One example is the open-source \ac{vcml}~\cite{machineware_machineware-gmbhvcml_2024}.

While SystemC-based simulations are widely used, alternative solutions, such as \ac{qemu}~\cite{bellard_qemu_2005}, exist.
\ac{qemu} is a \ac{cpu}-centric simulator supporting various host and target architectures.
Compared to frameworks like SystemC, it lacks standardization which makes it challenging to integrate new models.
Additionally, \ac{qemu} does not have a concept of simulation time which limits the detail level of the models.
To overcome these limitations, solutions exist that wrap the processor model of \ac{qemu} in a SystemC module.
An example of such a project is the open-source \ac{avp64}~\cite{junger_fast_2019,junger_armv8_2023}, which is used as a \ac{iss}-based reference system in this work.

\subsection{ARM-on-ARM Virtualization}

Traditional \ac{cpu} models use an \ac{iss} to translate instructions from the target \ac{isa} to the host \ac{isa}~\cite{bellard_qemu_2005,junger_sim-v_2022,magnusson_simics_2002,binkert_gem5_2011}.
Other approaches use recompilation of the target \ac{sw} to the host \ac{isa} followed by native execution~\cite{gerstlauer_host-compiled_2010,shen_native_2012}.
While these approaches avoid the overhead of an \ac{iss}, they come with limitations such as required source-code access of the target \ac{sw}.

A \ac{cpu} model needs to execute the instructions of the target \ac{sw}, simulate the \ac{cpu} state including multiple \acp{el}, accept interrupts, and perform the address translations of the \ac{mmu}.
Virtualization extensions allow to perform all these steps in \ac{hw} on the host without requiring a complex \ac{sw} solution.
They add a layer of abstraction between the \ac{hw} and the \ac{os}.
A hypervisor running in this layer can allow so-called guests to run side-by-side with the main \ac{os} (see \cref{fig:kvm}).
Linux offers the hypervisor \ac{kvm}~\cite{dall_kvmarm_2014} to use these virtualization features.

In 2020, the authors of~\cite{junger_arm--arm_2020} demonstrated how Linux's \ac{kvm} can be used to build a SystemC-based \ac{cpu} model that executes the target \ac{sw} natively on an ARM host without needing an \ac{iss}.
The work showed a basic proof of concept of a single-core \ac{vp}.
Their \ac{vp} reached speedups of up to \SI{2.57}{x} compared to an \ac{iss}-based \ac{vp}, although the ARM \ac{hw} they used was less performant compared to the x86 machine running the \ac{iss}.

In summary, \acp{vp} offer numerous advantages \ac{sw} development.
High simulation performance is essential for efficient development and testing.
Previous work showed that using Linux's \ac{kvm} for SystemC-based \ac{cpu} models results in huge speedups compared to traditional \ac{iss}-based models.
However, the implementation relied on additional \ac{hw} features and the possibility of applying custom kernel patches.
In this work, we present how to remove these limitations and extend the \ac{cpu} model to a multicore model.


\section{Approach}
\label{sec:approach}

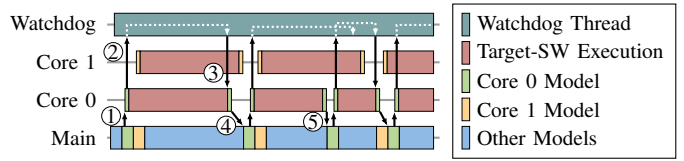
\begin{figure}[!t]
    \centering
    \tikzstyle{lbl} = [font={\footnotesize}, inner sep=1pt]
\tikzstyle{numbernode} = [font={\footnotesize}, inner sep=0pt, circle, draw=black, fill=white, inner sep=0.1pt, pos=0.7, line width=0.1mm]
\tikzstyle{basicnode} = [rectangle, minimum height=0.3cm, anchor=west, fill=rwth-50, inner sep=0pt, draw=black, minimum width=0.15cm, font={\color{white}\scriptsize}, line width=0.1mm, xshift=-0.1mm]

\pgfdeclarelayer{background}
\pgfdeclarelayer{middle}
\pgfdeclarelayer{front}
\pgfsetlayers{background,middle,main,front}

\def\vdist{0.5cm}
\def\vdistleg{0.375cm} 

\begin{tikzpicture}

    \node[lbl] (th_sc) {Main};
    \node[lbl, anchor=east] at ([yshift=\vdist]th_sc.east) (th_cpu0) {Core 0};
    \node[lbl, anchor=east] at ([yshift=\vdist]th_cpu0.east) (th_cpu1) {Core 1};
    \node[lbl, anchor=east] at ([yshift=\vdist]th_cpu1.east) (th_wd) {Watchdog};

    \phantom{\node[inner sep=0pt] at ([xshift=0.1cm]th_sc.east) (th_begin) {};}
    \phantom{\node[inner sep=0pt] at ([xshift=4.35cm]th_begin.east) (th_end) {};}

    \draw[black-50, thick] (th_sc.east-|th_begin) -- (th_sc.east-|th_end);
    \draw[black-50, thick] (th_cpu0.east-|th_begin) -- (th_cpu0.east-|th_end);
    \draw[black-50, thick] (th_cpu1.east-|th_begin) -- (th_cpu1.east-|th_end);
    \draw[black-50, thick] (th_wd.east-|th_begin) -- (th_wd.east-|th_end);

    \node[basicnode, fill=petrol-50, minimum width=4.20cm] at ([xshift=0.1cm]th_wd.east-|th_begin) (wd_runner) {};

    \node[basicnode, fill=rwth-50, minimum width=0.15cm] at ([xshift=0.05cm]th_sc.east-|th_begin) (mod0) {};
    \node[basicnode, fill=grun-50, minimum width=0.15cm] at (mod0.east) (mod_cpu0) {};
    \node[basicnode, fill=orange-50, minimum width=0.15cm] at (mod_cpu0.east) (mod_cpu1) {};
    \node[basicnode, fill=rwth-50, minimum width=1.3cm] at (mod_cpu1.east) (mod1) {};
    \node[basicnode, fill=grun-50, minimum width=0.15cm] at (mod1.east) (mod_cpu2) {};
    \node[basicnode, fill=orange-50, minimum width=0.15cm] at (mod_cpu2.east) (mod_cpu3) {};
    \node[basicnode, fill=rwth-50, minimum width=0.8cm] at (mod_cpu3.east) (mod2) {};
    \node[basicnode, fill=grun-50, minimum width=0.15cm] at (mod2.east) (mod_cpu4) {};
    \node[basicnode, fill=rwth-50, minimum width=0.5cm] at (mod_cpu4.east) (mod3) {};
    \node[basicnode, fill=orange-50, minimum width=0.15cm] at (mod3.east) (mod_cpu5) {};
    \node[basicnode, fill=grun-50, minimum width=0.15cm] at (mod_cpu5.east) (mod_cpu6) {};
    \node[basicnode, fill=rwth-50, minimum width=0.45cm] at (mod_cpu6.east) (mod4) {};

    \node[basicnode, fill=grun-50, minimum width=0.05cm] at ([xshift=0.05cm]mod_cpu0.west|-th_cpu0) (kvm0_setup) {};
    \node[basicnode, fill=bordeaux-50, minimum width=1.3cm] at (kvm0_setup.east) (kvm0) {};
    \node[basicnode, fill=grun-50, minimum width=0.05cm] at (kvm0.east) (kvm0_end) {};

    \begin{pgfonlayer}{front}
        \draw[-{Latex[length=1mm]}, thick] (kvm0_setup.west|-mod_cpu0.north) to node[numbernode, left, xshift=-0.05cm] {1} (kvm0_setup.south west);
        \draw[-{Latex[length=1mm]}, thick] (kvm0_setup.north) to node[numbernode, left, xshift=-0.05cm] {2} (kvm0_setup.north|-wd_runner.south);
        \draw[-{Latex[length=1mm]}, thick] (wd_runner.south-|kvm0_end.west) to node[numbernode, left, xshift=-0.05cm] {3} (kvm0_end.north west);
        \draw[-{Latex[length=1mm]}, thick] (kvm0_end.south east) to node[numbernode, left, xshift=-0.05cm, yshift=-0.05cm] {4} (mod_cpu2.north west);

        \draw[-{Latex[length=1mm]}, thick, densely dotted, white] (kvm0_setup.north|-wd_runner.south) -- ++(0,0.15cm) -| (wd_runner.south-|kvm0_end.west);
    \end{pgfonlayer}

    \node[basicnode, fill=orange-50, minimum width=0.05cm] at ([xshift=0.05cm]mod_cpu1.west|-th_cpu1) (kvm1_setup) {};
    \node[basicnode, fill=bordeaux-50, minimum width=1.3cm] at (kvm1_setup.east) (kvm1) {};
    \node[basicnode, fill=orange-50, minimum width=0.05cm] at (kvm1.east) (kvm1_end) {};

    \node[basicnode, fill=grun-50, minimum width=0.05cm] at ([xshift=0.1cm]mod_cpu2.west|-th_cpu0) (kvm2_setup) {};
    \node[basicnode, fill=bordeaux-50, minimum width=0.9cm] at (kvm2_setup.east) (kvm2) {};
    \node[basicnode, fill=grun-50, minimum width=0.05cm] at (kvm2.east) (kvm2_end) {};

    \begin{pgfonlayer}{front}
        \draw[-{Latex[length=1mm]}, thick] (kvm2_setup.west|-mod_cpu2.north) to (kvm2_setup.south west);
        \draw[-{Latex[length=1mm]}, thick] (kvm2_setup.north) to (kvm2_setup.north|-wd_runner.south);
        \draw[-{Latex[length=1mm]}, thick] (kvm2_end.south east) to node[numbernode, left, xshift=-0.05cm] {5} (mod_cpu4.north west);

        \draw[-{Latex[length=1mm]}, thick, densely dotted, white] (kvm2_setup.north|-wd_runner.south) -- ++(0,0.125cm) -| ([xshift=1.325cm]kvm2_setup.north|-wd_runner.south);
    \end{pgfonlayer}

    \node[basicnode, fill=orange-50, minimum width=0.05cm] at ([xshift=0.05cm]mod_cpu3.west|-th_cpu1) (kvm3_setup) {};
    \node[basicnode, fill=bordeaux-50, minimum width=1.3cm] at (kvm3_setup.east) (kvm3) {};
    \node[basicnode, fill=orange-50, minimum width=0.05cm] at (kvm3.east) (kvm3_end) {};

    \node[basicnode, fill=grun-50, minimum width=0.05cm] at ([xshift=0.1cm]mod_cpu4.west|-th_cpu0) (kvm4_setup) {};
    \node[basicnode, fill=bordeaux-50, minimum width=0.5cm] at (kvm4_setup.east) (kvm4) {};
    \node[basicnode, fill=grun-50, minimum width=0.05cm] at (kvm4.east) (kvm4_end) {};

    \begin{pgfonlayer}{front}
        \draw[-{Latex[length=1mm]}, thick] (kvm4_setup.west|-mod_cpu4.north) to (kvm4_setup.south west);
        \draw[-{Latex[length=1mm]}, thick] (kvm4_setup.north) to (kvm4_setup.north|-wd_runner.south);
        \draw[-{Latex[length=1mm]}, thick] (wd_runner.south-|kvm4_end.west) to (kvm4_end.north west);
        \draw[-{Latex[length=1mm]}, thick] (kvm4_end.south east) to (mod_cpu6.north west);

        \draw[-{Latex[length=1mm]}, thick, densely dotted, white] (kvm4_setup.north|-wd_runner.south) -- ++(0,0.175cm) -| (wd_runner.south-|kvm4_end.west);
    \end{pgfonlayer}

    \node[basicnode, fill=grun-50, minimum width=0.05cm] at ([xshift=0.1cm]mod_cpu6.west|-th_cpu0) (kvm6_setup) {};
    \node[basicnode, fill=bordeaux-50, minimum width=0.46cm] at (kvm6_setup.east) (kvm6) {};

    \begin{pgfonlayer}{front}
        \draw[-{Latex[length=1mm]}, thick] (kvm6_setup.west|-mod_cpu6.north) to (kvm6_setup.south west);
        \draw[-{Latex[length=1mm]}, thick] (kvm6_setup.north) to (kvm6_setup.north|-wd_runner.south);

        \draw[densely dotted, thick, white] (kvm6_setup.north|-wd_runner.south) |- ++(0.48cm,0.15cm);
    \end{pgfonlayer}

    \node[basicnode, fill=orange-50, minimum width=0.05cm] at ([xshift=0.1cm]mod_cpu5.west|-th_cpu1) (kvm5_setup) {};
    \node[basicnode, fill=bordeaux-50, minimum width=0.61cm] at (kvm5_setup.east) (kvm5) {};

    \node[basicnode, xshift=0pt, fill=petrol-50, minimum width=0.1cm] at ([xshift=0.3cm]th_wd-|th_end) (leg_wd) {};
    \node[basicnode, xshift=0.1mm, fill=bordeaux-50, minimum width=0.1cm, anchor=center] at ([yshift=-\vdistleg]leg_wd.center) (leg_kvm) {};
    \node[basicnode, xshift=0.1mm, fill=grun-50, minimum width=0.1cm, anchor=center] at ([yshift=-\vdistleg]leg_kvm.center) (leg_cpu0) {};
    \node[basicnode, xshift=0.1mm, fill=orange-50, minimum width=0.1cm, anchor=center] at ([yshift=-\vdistleg]leg_cpu0.center) (leg_cpu1) {};
    \node[basicnode, xshift=0.1mm, fill=rwth-50, minimum width=0.1cm, anchor=center] at ([yshift=-\vdistleg]leg_cpu1.center) (leg_sc) {};

    \node[lbl, anchor=west] at ([xshift=0.05cm]leg_wd.east) (leg_wd_txt) {Watchdog Thread};
    \node[lbl, anchor=west] at (leg_wd_txt.west|-leg_kvm) (leg_kvm_txt) {Target-\acs{sw} Execution};
    \node[lbl, anchor=west] at (leg_wd_txt.west|-leg_cpu0) (leg_cpu0_txt) {Core 0 Model};
    \node[lbl, anchor=west] at (leg_wd_txt.west|-leg_cpu1) (leg_cpu1_txt) {Core 1 Model};
    \node[lbl, anchor=west] at (leg_wd_txt.west|-leg_sc) (leg_sc_txt) {Other Models};

    \begin{pgfonlayer}{background}
        \node[fit={(leg_wd) (leg_kvm_txt) (leg_sc)}, fill=white, draw=black] (legend) {};
    \end{pgfonlayer}

\end{tikzpicture}
    \caption{Multithreaded approach using a \ac{sw}-based watchdog.}
    \label{fig:approach}
\end{figure}

Our approach is sketched in \cref{fig:approach} for a dual-core setup.
In a traditional SystemC-based simulation, only the \textit{main thread} exists, where different models are simulated sequentially.
For our parallelized approach, the SystemC kernel and all models except our \ac{cpu} model run in the main thread.
The \ac{cpu} model has a placeholder in the main thread that is used for synchronization and the communication with other models.
For the actual work, a separate thread is used.
Details of this parallelization technique can be found in~\cite{bosbach_towards_2024,bosbach_work--progress_2023}.

For fast \acp{fss}, the \textit{loosely-timed} coding style is usually used to abstract timing for increased performance~\cite{systemc-2023}.
A technique used with this coding style is \textit{temporal decoupling}.
It permits SystemC processes to run ahead of the global simulation time before they need to synchronize again.
A parameter called \textit{quantum} defines how far a process is allowed to run ahead to control the trade-off between performance and accuracy.
Temporal decoupling allows the \ac{cpu} model to execute an entire quantum of instructions before synchronization is needed.

When the placeholder \textit{SC\_THREAD} of a \ac{cpu} model is scheduled by SystemC, instructions of the target \ac{sw} can be simulated.
The worker running in the asynchronous thread, e.g., \textit{Core 0} in \cref{fig:approach}, is informed to execute a defined number of instructions~\circled{1}.
While other models continue their simulation in the main thread, a \ac{sw}-based watchdog running in a different thread is programmed to stop the execution of target \ac{sw} after the specified wall-clock-time interval~\circled{2}.
Then, the target \ac{sw} is executed natively on the host using the virtualization features.
When the watchdog expires~\circled{3}, the execution of the target \ac{sw} is suspended and the placeholder \textit{SC\_THREAD} running in the main thread is notified~\circled{4}.

Depending on the instructions of the target \ac{sw}, the execution might be suspended before watchdog expiration.
This is, e.g., the case when the target \ac{sw} needs to access a memory-mapped region of a peripheral.
Then, the access to the peripheral is performed in the main thread~\circled{5}.
Afterwards, the execution of the target \ac{sw} can be continued.
In case of an early exit, the programmed watchdog is not needed (details in \cref{sec:implementation:watchdog}).

In this work, we present a technique to force early accesses when the Linux of the target \ac{sw} executes an idle loop.
The Linux idle loop uses the \ac{wfi} instruction to hint a possible suspension until an interrupt is signaled.
We can use this forced early access to skip the execution of idle loops and thereby speed up the simulation.
Details can be found in \cref{sec:implementation:wfi}.

\section{Implementation}
\label{sec:implementation}

We build our implementation of the \ac{kvm}-based processor model on top of the SystemC-\ac{tlm}-based, open-source \ac{vcml}~\cite{machineware_machineware-gmbhvcml_2024} project.
\ac{vcml} offers the \lstinline{processor} class as a starting point to integrate an instruction interpreter, simulator, or executor into a loosely-timed SystemC module.
It implements a basic simulation loop that calls the \lstinline{virtual void simulate(size_t cycles)} function to simulate the specified number of instructions.
This function is used to call either the \ac{iss} or, in our case, \ac{kvm}.
Additionally, parallel execution of the simulate function can be activated to offload the execution to a separate thread on the host machine~\cite {bosbach_work--progress_2023,bosbach_towards_2024}.

Communication with \ac{kvm} is mainly done through system calls.
To efficiently handle memory accesses of the target \ac{sw}, a memory region from the \ac{vp} process, can be mapped to the \ac{kvm} guest.
We use this feature to map the virtual memory model of the \ac{vp} to the guest environment.
The region is queried using \ac{tlm}-\ac{dmi}, similar to how an \ac{iss} would operate.
This allows the native execution of load and store instructions to memory.

\subsection{Working Principle}
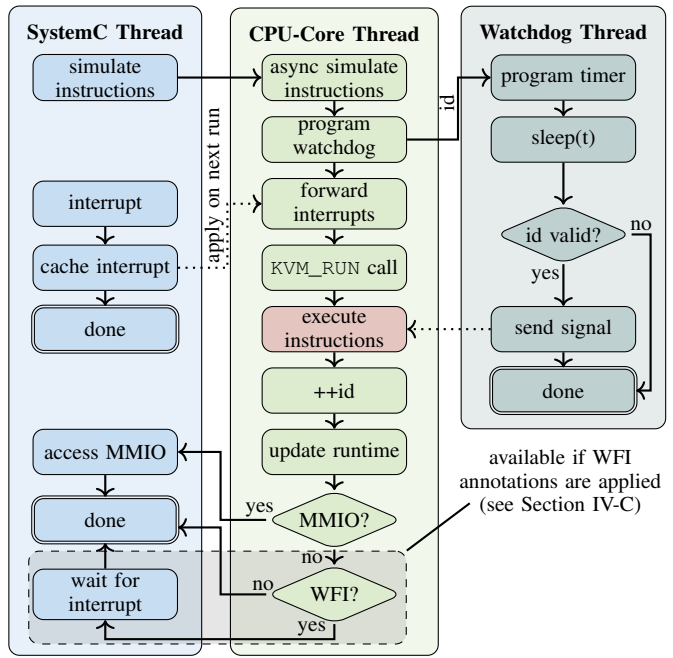
\begin{figure}[!t]
    \centering
    \tikzstyle{nodetype} = [rectangle, rounded corners, minimum width=1.8cm, minimum height=.6cm, text centered, text width=1.8cm, draw=black, fill=rwth-25, font={\footnotesize}, inner sep=0.05cm]
\tikzstyle{diamondtype} = [nodetype, diamond, minimum height=0.6cm, minimum width=1.0cm, text width=1.0cm, aspect=2.3]
\tikzstyle{lbl}      = [font={\footnotesize}, inner sep=1pt]

\pgfdeclarelayer{background}
\pgfdeclarelayer{middle}
\pgfsetlayers{background,middle,main}

\begin{tikzpicture}
    \node[nodetype] (sim_instr) {simulate instructions};
    \node[nodetype, fill=grun-25, anchor=west] at ([xshift=1.1cm]sim_instr.east) (async_sim_instr) {async simulate instructions};
    \node[nodetype, fill=grun-25, anchor=north] at ([yshift=-0.2cm]async_sim_instr.south) (start_wd) {program watchdog};
    \node[nodetype, fill=grun-25, anchor=north] at ([yshift=-0.2cm]start_wd.south) (update_irq) {forward interrupts};
    \node[nodetype, fill=grun-25, anchor=north] at ([yshift=-0.2cm]update_irq.south) (kvm_run) {\texttt{KVM\_RUN} call};
    \node[nodetype, fill=bordeaux-25, anchor=north] at ([yshift=-0.2cm]kvm_run.south) (exec) {execute instructions};
    \node[nodetype, fill=grun-25, anchor=north] at ([yshift=-0.2cm]exec.south) (exit) {++id};
    \node[nodetype, fill=grun-25, anchor=north] at ([yshift=-0.2cm]exit.south) (update_time) {update runtime};
    \node[diamondtype, fill=grun-25, anchor=north] at ([yshift=-0.2cm]update_time.south) (mmio) {MMIO?};
    \node[diamondtype, fill=grun-25, anchor=north] at ([yshift=-0.2cm]mmio.south) (wfi) {WFI?};
    \node[nodetype, anchor=center] at (sim_instr|-update_time) (access_mmio) {access MMIO};
    \node[nodetype, anchor=center, double=rwth-10] at (access_mmio|-mmio) (done) {done};
    \node[nodetype, anchor=center] at (sim_instr|-wfi) (wfi_sc) {wait for interrupt};

    \node[nodetype, fill=petrol-25, anchor=west] at ([xshift=1.1cm]async_sim_instr.east) (wd_setup) {program timer};
    \node[nodetype, fill=petrol-25, anchor=center] at (wd_setup|-start_wd) (sleep) {sleep(t)};
    \phantom{\node at ($(update_irq.north)!0.5!(kvm_run.south)$)  (helper_id_check) {};}
    \node[diamondtype, fill=petrol-25, anchor=center] at (sleep|-helper_id_check) (id_check) {id valid?};
    \node[nodetype, fill=petrol-25, anchor=center] at (id_check|-exec) (kick) {send signal};
    \node[nodetype, fill=petrol-25, anchor=center, double=petrol-10] at (kick|-exit) (wd_done) {done};

    \node[nodetype, anchor=center] at (sim_instr|-update_irq) (irq) {interrupt};
    \node[nodetype, anchor=center] at (irq|-kvm_run) (irq_local_update) {cache interrupt};
    \node[nodetype, anchor=center, double=rwth-10] at (irq_local_update|-exec) (irq_done) {done};

    \draw[->, thick] (sim_instr.east) -- (async_sim_instr.west);
    \draw[->, thick] (async_sim_instr.south) -- (start_wd.north);
    \draw[->, thick] (start_wd.south) -- (update_irq.north);
    \draw[->, thick] (update_irq.south) -- (kvm_run.north);
    \draw[->, thick] (kvm_run.south) -- (exec.north);
    \draw[->, thick] (exec.south) -- (exit.north);
    \draw[->, thick] (exit.south) -- (update_time.north);
    \draw[->, thick] (update_time.south) -- (mmio.north);
    \draw[->, thick] (mmio.west) -- ++(-0.6cm,0cm) node[lbl, above, pos=0, anchor=south east, xshift=0.15cm] {yes} |- (access_mmio.east);
    \draw[->, thick] (mmio.south) -- node[lbl, anchor=east, xshift=-0.1cm, yshift=-0.5pt] {no} (wfi.north);
    \draw[->, thick] (access_mmio.south) -- (done.north);
    \draw[->, thick] (wfi.west) -- ++(-0.6cm,0cm) node[lbl, above, pos=0, anchor=south east, xshift=0.15cm] {no} |- ([yshift=-0.1cm]done.east);
    \draw[->, thick] (wfi.south) -- ++(0cm,-0.2cm) -| node[lbl, above, pos=0.01, anchor=south east] {yes} (wfi_sc.south);
    \draw[->, thick] (wfi_sc.north) -- (done.south);

    \draw[->, thick] (irq.south) -- (irq_local_update.north);
    \draw[->, thick] (irq_local_update.south) -- (irq_done.north);

    \draw[->, thick] (start_wd.east) -- ++(0.67cm, 0cm) |- node[lbl, pos=0.2, anchor=south west, rotate=90] {id} (wd_setup.west);
    \draw[->, thick] (wd_setup.south) -- (sleep.north);
    \draw[->, thick] (sleep.south) -- (id_check.north);
    \draw[->, thick] (id_check.south) -- node[lbl, anchor=east, pos=0.3] {yes} (kick.north);
    \draw[->, thick, dotted] (kick.west) -- (exec.east);
    \draw[->, thick] (kick.south) -- (wd_done.north);
    \draw[->, thick] (id_check.east) -- node[lbl, anchor=south] {no}  ++(0.25cm, 0cm) |- (wd_done.east);

    \draw[->, thick, dotted] (irq_local_update.east) -- ++(0.65cm, 0cm) |- node[lbl, pos=0.01, rotate=90, above, anchor=south west] {apply on next run} (update_irq.west);

    \node[lbl, anchor=south] at ([yshift=0.1cm]sim_instr.north) (sc_thread) {\textbf{SystemC Thread}};
    \node[lbl, anchor=center] at (sc_thread-|async_sim_instr) (async_thread) {\textbf{CPU-Core Thread}};
    \node[lbl, anchor=center] at (sc_thread-|wd_setup) (wd_thread) {\textbf{Watchdog Thread}};

    \begin{pgfonlayer}{background}
        \phantom{\node[lbl, inner sep=0pt] at (current bounding box.south) (bottom) {};}

        \node[lbl, anchor=south] at (bottom-|sim_instr) (dummy_sc_thread_bottom) {};
        \node[lbl, anchor=south] at (bottom-|async_sim_instr) (dummy_async_thread_bottom) {};
        \node[lbl, anchor=south] at (wd_done.south-|wd_setup) (dummy_wd_thread_bottom) {};

        \node[nodetype, fill=rwth-10, fit={(sc_thread) (done) (dummy_sc_thread_bottom)}, inner sep=0.2cm] (sc_thread_box) {};
        \node[nodetype, fill=grun-10, fit={(async_thread) (async_sim_instr) (dummy_async_thread_bottom)}, inner sep=0.2cm] (cpu_thread_box) {};
        \node[nodetype, fill=petrol-10, fit={(wd_thread) (sleep) (dummy_wd_thread_bottom)}, inner sep=0.2cm] (wd_thread_box) {};
    \end{pgfonlayer}

    \begin{pgfonlayer}{middle}
        \node[nodetype, dashed, fit={([yshift=-2pt]mmio.south) (wfi) (wfi_sc) (wfi|-bottom)}, fill=black, fill opacity=0.1] (wfi_box) {};
    \end{pgfonlayer}

    \node[lbl, anchor=east, text width=2.7cm, text centered] at ([yshift=0.5cm]mmio-|wd_thread_box.east) (wfi_box_text) {available if \acs{wfi} annotations are applied (see \cref{sec:implementation:wfi})};
    \draw[thick] ([xshift=0.2cm, yshift=0.2cm]wfi_box_text.south west) -- (wfi_box.north east);

\end{tikzpicture}
    \vspace{1em}
    \caption{\ac{vcpu} model execution loop.}
    \label{fig:loop}
\end{figure}

A basic overview of the \ac{cpu} model's simulate function is given in \cref{fig:loop}.
When parallel execution is activated, the call is executed in a separate thread for each simulated \ac{cpu} core to allow other models of the simulation to run in parallel.
Otherwise, the \textit{\ac{cpu}-Core Thread} does not exist, and the functions are directly executed in the SystemC thread.

First, the allowed runtime of \ac{kvm} is calculated from the passed number of cycles.
As usual for instruction-accurate simulators, a constant average execution time per instruction is assumed.
The clock frequency is obtained from the virtual clock that is connected to the SystemC processor module.
The watchdog is programmed to expire after the calculated amount of time (more details on the watchdog follow in \cref{sec:implementation:watchdog}).

Pending interrupts are injected.
The (wall-clock) timestamp is stored in a local variable and the execution of \ac{kvm} is started using the \texttt{KVM\_RUN} call.
\ac{kvm} then executes the guest code.
Certain events are trapped by \ac{kvm} that lead to a return to user space and therefore an exit of the \texttt{KVM\_RUN} call.
The relevant events for this work are \ac{mmio} accesses, breakpoint hits, and received Linux signals.

After an exit of \texttt{KVM\_RUN}, the internal watchdog \ac{id} is incremented (more on this in \cref{sec:implementation:watchdog}), and the duration of the \texttt{KVM\_RUN} call is determined by subtracting the current (wall-clock) timestamp from the previously stored one.
This runtime is used as an approximation for the number of executed instructions and thereby cycles.

Then, the cause of the \texttt{KVM\_RUN} exit is determined.
In case of an \ac{mmio} access, a  \ac{tlm} transaction is sent to the corresponding peripheral.
Since, according to the SystemC standard, interactions with other models need to be performed in the main thread, this access has to be shifted back in case of parallel execution~\cite{systemc-2023,bosbach_towards_2024}.
If the exit was caused by an idling core that executed a \ac{wfi} instruction, performance optimization can be applied (more details in \cref{sec:implementation:wfi}).

Once the \ac{mmio} handling is done, the \ac{wfi} hint has been processed, or if the exit has been caused by the watchdog, the \lstinline{simulate} function returns.
The loop of the \ac{vcml} processor class will continue and call the \lstinline{simulate} function again.

\subsection{Software-Based Watchdog Timer}
\label{sec:implementation:watchdog}

To stop the execution after the end of a quantum, a \ac{sw}-based watchdog timer is used (see \cref{sec:approach}).
The \ac{vcpu} clock is used to convert the quantum into the number of instructions that can be executed before synchronizing.
Once the number of instructions has been executed, a Linux signal is sent to the thread that executes the \texttt{KVM\_RUN} call to stop the execution.


Recent approaches have used Linux's \ac{cpu}-performance-counter-API \textit{perf} to count the executed instructions in guest mode and send a signal once a threshold is exceeded~\cite{junger_arm--arm_2020}.
While this method provides high accuracy, it depends on specific \ac{hw} features that may not be universally available.
For instance, Apple's \acp{cpu} contain custom \acp{pmu} instead of the standard ARM one.
In Asahi Linux~\cite{asahi_linux_asahi_nodate}, perf cannot be configured to send a signal once a specified amount of guest instructions has been executed.
Therefore, we use an approach that does not rely on \ac{hw}-specific features.

Before invoking \texttt{KVM\_RUN}, we set up the \ac{sw}-based watchdog timer.
The watchdog timer is shared between all cores and runs in a separate thread (see \cref{fig:approach,fig:loop}).
It calls the \lstinline{kick} function after a predefined timeout that is calculated by dividing the maximum number of instructions to be executed by the \ac{vcpu} clock frequency.
The kick function and the scheduling of the timer are shown in \cref{lst:kick}.

\begin{lstlisting}[label=lst:kick, float, caption={Kick function that is called by the watchdog.}, escapechar=|]
// kick KVM
void cpu::kick(unsigned int id) {
    if (id == m_kickid) |\label{line:kick:checkid}|
        pthread_kill(m_self, SIGUSR1); |\label{line:kick:signal}|
}
// schedule watchdog
watchdog(timeout, [&, id = m_kickid]() -> void { |\label{line:kick:passid}|
            kick(id); }); |\label{line:kick:passid-end}|
\end{lstlisting}

Our \ac{cpu} class maintains an internal \ac{id} counter (\lstinline{m_kickid}) that is incremented after each \texttt{KVM\_RUN} to identify a run.
When scheduling the watchdog timer, we pass this current \ac{id} value (\cref{line:kick:passid,line:kick:passid-end}).
Upon expiration, the \ac{id} is used to only send the \texttt{SIGUSR1} signal to exit \ac{kvm} if the corresponding \texttt{KVM\_RUN} is still active (\cref{line:kick:checkid,line:kick:signal}).
This approach effectively limits the maximum \ac{kvm} run time.

By employing this method, we achieve robust synchronization without relying on hardware-specific performance counters.
Consequently, our solution is versatile and compatible with various environments, such as Asahi Linux on Apple Silicon.

\subsection{WFI Annotations}
\label{sec:implementation:wfi}

The ARM architecture includes a \ac{wfi} instruction, that is commonly used in idle loops within \acp{os} to suspend \ac{cpu} activity until an interrupt occurs~\cite{arm_ltd_arm_2024}.
In a SystemC-based, event-driven simulation, pausing the execution of the \ac{cpu} model during idle periods and resuming it upon receiving an interrupt can significantly enhance simulation performance by skipping idle time instead of simulating it.

However, \ac{kvm} does not inherently notify the process that started a \ac{kvm} guest about executed \ac{wfi} instructions.
The \ac{wfi} instructions are trapped by \ac{kvm} and Linux's scheduler is called to schedule another user thread.
Therefore, modifications in \ac{kvm} or workarounds are necessary to get notified of \ac{wfi} instruction execution to then use SystemC's feature to pause the model until the next interrupt occurs.

Previous approaches used custom patches to send notifications to user space~\cite{junger_arm--arm_2020}.
While this is an effective approach for research purposes, this method comes with several drawbacks:

\begin{itemize}
    \item \textit{Security concerns}: Stringent security policies may prevent companies from applying custom patches.
    \item \textit{Maintenance overhead}: Custom patches need reapplication with every kernel update, complicating maintenance.
    \item \textit{Mainline-integration challenges}: Efforts to integrate such patches into the mainline Linux kernel failed. 
\end{itemize}

To address these limitations, we propose an alternative way to trap \ac{wfi} instructions without requiring kernel patches.
This approach works when a Linux \ac{os} is executed on the target system since Linux uses the \ac{wfi} instruction only in its idle loop.
Our solution involves searching for specific symbols within the target \ac{sw}'s \ac{elf} file during \ac{vp} startup and setting breakpoints accordingly.
The steps are as follows:

\begin{enumerate}
    \item \textit{Symbol search}: Identify the \lstinline{cpu_do_idle} symbol within the target-\ac{sw} \ac{elf} file.
    \item \textit{Breakpoint setting}: Locate the \ac{wfi} instruction inside this function and set a breakpoint on it.
    \item \textit{Instruction check}: When a \ac{vcpu} reaches this breakpoint, \ac{kvm} exits back to user space.
    \item \textit{Program counter verification}: Verify if the program counter matches the address of the \ac{wfi} instruction to distinguish it from other breakpoints set by users.
\end{enumerate}

Upon confirming that a \ac{cpu} intends to execute a \ac{wfi} instruction, we use SystemC's features to suspend model execution until an interrupt is signaled.
This technique avoids the simulation of idle loops, thereby improving the overall simulation efficiency and performance.

We refer to this method as \textit{\ac{wfi} annotation}.
It provides a robust way to handle idle states in multicore simulations without compromising security or maintainability.

\section{Results}
\label{sec:results}

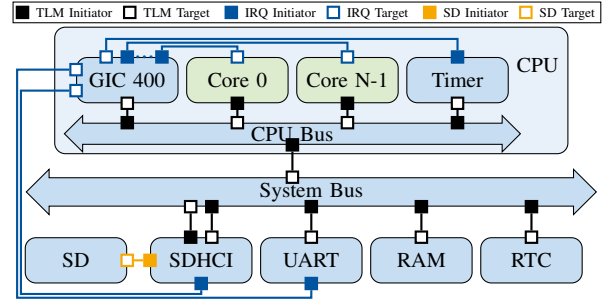
\begin{figure}
    \centering
    \tikzstyle{nodetype} = [rectangle, rounded corners, minimum width=1.1cm, minimum height=.6cm, text centered, text width=1.1cm, draw=black, fill=rwth-25, font={\footnotesize}]
\tikzstyle{isock}    = [rectangle, minimum width=.15cm, minimum height=.15cm, draw=black, fill=black, inner sep=0pt, thick]
\tikzstyle{tsock}    = [isock, fill=white]
\tikzstyle{isocki}   = [isock, fill=rwth, draw=rwth]
\tikzstyle{tsocki}   = [isocki, fill=white]
\tikzstyle{isockp}   = [isock, fill=grun, draw=grun]
\tikzstyle{tsockp}   = [isockp, fill=white]
\tikzstyle{isocksd}   = [isock, fill=orange, draw=orange]
\tikzstyle{tsocksd}   = [isocksd, fill=white]
\tikzstyle{lbl}      = [font={\tiny}, inner sep=1pt, text depth=0pt]
\tikzstyle{socklbl}  = [lbl]

\pgfdeclarelayer{background}
\pgfsetlayers{background,main}

\begin{tikzpicture}[node distance=.5cm and .5cm]

    \begin{scope}[name prefix=core0-]
        \node[nodetype, fill=grun-25, anchor=west] (main) {Core 0};
        \node[isock] at (main.south)(sock) {};
        \node[tsocki] at (main.north)(socki) {};
    \end{scope}

    \begin{scope}[name prefix=coren1-]
        \node[nodetype, fill=grun-25, anchor=west] (main) at ([xshift=0.1cm]core0-main.east) {Core N-1};
        \node[isock] at (main.south)(sock) {};
        \node[tsocki] at (main.north)(socki) {};
    \end{scope}

    \begin{scope}[name prefix=timer-]
        \node[nodetype, anchor=west] (main) at ([xshift=0.1cm]coren1-main.east) {Timer};
        \node[tsock] at (main.south)(sock) {};
        \node[isocki] at (main.north)(socki) {};
    \end{scope}

    \begin{scope}[name prefix=gic-]
        \node[nodetype, anchor=east] (main) at ([xshift=-0.1cm]core0-main.west) {GIC 400};

        \node[tsock, xshift=+0.0cm] at (main.south) (sock) {};

        \node[isocki] at (main.north) (sock_intin1) {};
        \node[yshift=+0.0cm, text=rwth, anchor=west, inner sep=0] at (sock_intin1.east) (sock_inti) {...};
        \node[isocki, anchor=west] at (sock_inti.east) (sock_inti0) {};
        \node[tsocki, anchor=east] at ([xshift=-0.1cm]sock_intin1.west) (sock_int0) {};

        \node[tsocki, anchor=north] at ([yshift=-0.05cm]main.west) (sock_int1) {};
        \node[tsocki, anchor=south] at ([yshift=0.1cm]sock_int1.north) (sock_int2) {};
    \end{scope}

    \draw[thick, draw=rwth] (core0-socki.north) -- +(0,+0.05cm) -| (gic-sock_inti0.north);
    \draw[thick, draw=rwth] (coren1-socki.north) -- +(0,+0.1cm) -| (gic-sock_intin1.north);
    \draw[thick, draw=rwth] (timer-socki.north) -- +(0,+0.15cm) node[inner sep=0pt] (cpu-connection) {} -| (gic-sock_int0.north);

    \node[fit={(core0-main) (coren1-main) (gic-main) (timer-main) (cpu-connection)}, inner sep=0pt] (cpu-components) {};

    \begin{scope}[name prefix=corebus-]
        \node[double arrow, draw, minimum height=6cm, fill=rwth-25, font={\footnotesize}, anchor=north, inner sep=1pt,  double arrow head extend=0.1cm] at ([yshift=-0.25cm]cpu-components.south) (main) {CPU Bus};

        \node[isock] at (main.north-|gic-sock) (gic){};
        \draw[thick] (gic) -- (gic-sock);

        \node[tsock] at (main.north-|core0-sock) (core0){};
        \draw[thick] (core0) -- (core0-sock);

        \node[tsock] at (main.north-|coren1-sock) (coren1){};
        \draw[thick] (coren1) -- (coren1-sock);

        \node[isock] at (main.north-|timer-sock) (timer){};
        \draw[thick] (timer) -- (timer-sock);

        \node[isock] at (main.south) (sysbus){};
    \end{scope}

    \node[font={\footnotesize}, inner sep=0pt, anchor=north west] at ([xshift=0.1cm]timer-main.north east) (cpu-text) {CPU};
    \begin{pgfonlayer}{background}
        \node[nodetype, fit={(cpu-components) (corebus-main) (cpu-text)}, fill=rwth-10] (cpu-main) {};
    \end{pgfonlayer}

    \begin{scope}[name prefix=uart-]
        \node[nodetype, anchor=north] at ([yshift=-1.1cm]cpu-main.south) (main) {UART};
        \node[tsock] at (main.north)(sock) {};
        \node[isocki] at (main.south)(socki) {};
    \end{scope}

    \begin{scope}[name prefix=ram-]
        \node[nodetype, anchor=west] at ([xshift=.1cm]uart-main.east) (main) {RAM};
        \node[tsock] at (main.north)(sock) {};
    \end{scope}

    \begin{scope}[name prefix=rtc-]
        \node[nodetype, anchor=west] at ([xshift=0.1cm]ram-main.east) (main) {RTC};
        \node[tsock] at (main.north)(sock) {};
    \end{scope}

    \begin{scope}[name prefix=sdhci-]
        \node[nodetype, anchor=east] at ([xshift=-0.1cm]uart-main.west) (main) {SDHCI};
        \node[tsock, anchor=west] at ([xshift=0.05cm]main.north)(sock) {};
        \node[isock, anchor=east] at ([xshift=-0.05cm]main.north)(isock) {};
        \node[isocki] at (main.south)(socki) {};
        \node[isocksd] at (main.west)(socksd) {};
    \end{scope}

    \begin{scope}[name prefix=sd-]
        \node[nodetype, anchor=east] at ([xshift=-0.3cm]sdhci-main.west) (main) {SD};
        \node[tsocksd] at (main.east)(socksd) {};
    \end{scope}

    \begin{scope}[name prefix=bus-]
        \node[double arrow, draw, minimum height=7.5cm, minimum width=0.6cm, fill=rwth-25, font={\footnotesize}, anchor=center, inner sep=1pt,  double arrow head extend=0.1cm] at ([yshift=-0.5cm]cpu-main.south) (main) {System Bus};

        \node[tsock] at (main.north-|corebus-sysbus) (corebus){};
        \draw[thick] (corebus) -- (corebus-sysbus);

        \node[isock] at (main.south-|sdhci-sock) (sdhci){};
        \draw[thick] (sdhci) -- (sdhci-sock);
        \node[tsock] at (main.south-|sdhci-isock) (sdhcii){};
        \draw[thick] (sdhcii) -- (sdhci-isock);

        \node[isock] at (main.south-|ram-sock) (ram){};
        \draw[thick] (ram) -- (ram-sock);

        \node[isock] at (main.south-|rtc-sock) (rtc){};
        \draw[thick] (rtc) -- (rtc-sock);

        \node[isock] at (main.south-|uart-sock) (uart){};
        \draw[thick] (uart) -- (uart-sock);
    \end{scope}

    \draw[thick, draw=rwth] (sdhci-socki.south) -- ++(0cm, -0.05cm) -| ([xshift=-0.05cm]sd-main.west|-gic-sock_int1.west) -- (gic-sock_int1.west);
    \draw[thick, draw=rwth] (uart-socki.south) -- ++(0cm, -0.1cm) -| ([xshift=-0.1cm]sd-main.west|-gic-sock_int2.west) -- (gic-sock_int2.west);

    \draw[thick, draw=orange] (sdhci-socksd) -- (sd-socksd);

    \begin{scope}[name prefix=legend-]
        \node[isock, anchor=south] at ([xshift=-3.8cm,yshift=0.1cm]cpu-main.north) (isock) {};
        \node[lbl, anchor=west] at ([xshift=0.05cm]isock.east) (lbl_isock) {TLM Initiator};
        \node[tsock, anchor=west] at ([xshift=0.1cm]lbl_isock.east) (tsock) {};
        \node[lbl, anchor=west] at ([xshift=0.05cm]tsock.east) (lbl_tsock) {TLM Target};

        \node[isocki, anchor=west] at ([xshift=0.15cm]lbl_tsock.east) (isocki) {};
        \node[lbl, anchor=west] at ([xshift=0.05cm]isocki.east) (lbl_isocki) {IRQ Initiator};
        \node[tsocki, anchor=west] at ([xshift=0.1cm]lbl_isocki.east) (tsocki) {};
        \node[lbl, anchor=west] at ([xshift=0.05cm]tsocki.east) (lbl_tsocki) {IRQ Target};

        \node[isocksd, anchor=west] at ([xshift=0.15cm]lbl_tsocki.east) (isocksd) {};
        \node[lbl, anchor=west] at ([xshift=0.05cm]isocksd.east) (lbl_isocksd) {SD Initiator};
        \node[tsocksd, anchor=west] at ([xshift=0.1cm]lbl_isocksd.east) (tsocksd) {};
        \node[lbl, anchor=west] at ([xshift=0.05cm]tsocksd.east) (lbl_tsocksd) {SD Target};

        \begin{pgfonlayer}{background}
            \node[draw=black, fill=white, inner sep=1pt, fit={(isock) (lbl_tsocksd)}] {};
        \end{pgfonlayer}
    \end{scope}
\end{tikzpicture}
    \caption{\acs{aoa}-based \acs{vp}.}
    \label{fig:vp}
    \vspace{1em}
\end{figure}

To evaluate the performance results of our approach, we integrated the \ac{kvm}-based \ac{cpu} model into a \ac{vp}.
\Cref{fig:vp} shows an overview of the \ac{vp} architecture.
Each green-colored \ac{cpu} core launches a \ac{kvm}-based guest to execute target instructions.
The \ac{vcpu} consists of \num{1} to \num{8} \ac{kvm}-based cores, a \ac{gic}~400, and a memory-mapped timer.
The simulated peripherals include a \ac{sdhci} device with a virtual \ac{sd} card, an \ac{uart} interface for user interaction, \ac{ram}, and a \ac{rtc}.
All peripherals are taken from the open-source \ac{vcml} library~\cite{machineware_machineware-gmbhvcml_2024}.
Communication between the models is realized via \ac{tlm} sockets and protocols for memory accesses, interrupts, and \ac{sd}-card operations.

We conducted our experiments on a 10-core \textit{Apple Mac mini} equipped with an \textit{M2 Pro} processor, \SI{16}{\giga\byte} \ac{ram}, and \SI{512}{\giga\byte} \ac{ssd} memory.
The M2 Pro processor has \num{6} high-performance \textit{Avalanche} (\SI{3.7}{\giga\hertz}) cores and \num{4} efficiency \textit{Blizzard} (\SI{3.4}{\giga\hertz}) cores.
For comparison against traditional \ac{iss}-based \ac{cpu} models, we used the open-source \ac{avp64}~\cite{junger_fast_2019,junger_armv8_2023} running on an \textit{AMD Ryzen~9 3900X} (\SI{3.8}{\giga\hertz}, \SI{4.6}{\giga\hertz} boost) with \num{12} cores and \num{24} \ac{hw} threads.
Both \acp{vp} support a parallel execution mode that runs the \ac{cpu} cores in separate threads~\cite{bosbach_towards_2024,bosbach_work--progress_2023}.
Different quantum values are used to steer the temporal decoupling of the simulated models.

\subsection{Bare-Metal Dhrystone Benchmark}
\label{sec:results:dhrystone}

The first benchmark executed is a bare-metal Dhrystone~\cite{weicker_dhrystone_1988}, which is single threaded and integer based.
For multicore systems, each core executes its own instance of Dhrystone.
Thus, a parallel Dhrystone is an optimally parallelizable, compute-intensive workload that does not involve any communication.

\begin{figure}[!t]
    \centering

    \renewcommand{\uniquepgftag}{dhrystone}
    \pgfplotstableread[col sep=comma]{plot/data/aoavp_dhrystone_seq.csv}\resultsseq
    \pgfplotstableread[col sep=comma]{plot/data/aoavp_dhrystone_par_diff.csv}\resultsdiff
    \pgfplotstableread[col sep=comma]{plot/data/avp64_dhrystone_seq.csv}\refresultsseq
    \pgfplotstableread[col sep=comma]{plot/data/avp64_dhrystone_par_diff.csv}\refresultsdiff
    \pgfdeclarelayer{background}
\pgfsetlayers{background,main}

\begin{tikzpicture}
    \pgfplotsset{every axis/.style={ 
                BarConfig,
                bar width=4pt,
                ybar stacked,
                ymin=0,
                ymax=60000,
                minor y tick num=1,
                yminorgrids=true,
                major x tick style = {opacity=0},
                minor x tick num = 1,
                xticklabels from table={\resultsseq}{cores},
                x tick label style={yshift=-0.6cm, name=xlabel\ticknum},
                xticklabel style={rotate=0, anchor=north},
                ylabel={\shortstack{Accumulated\\MIPS}},
                xlabel={\# Simulated Cores \& Quantum},
                xlabel shift=-3pt,
                enlarge x limits=0.2,
                legend style={at={([yshift=-1pt]0.5,1.0)}, anchor=north, legend columns=4, inner sep=1pt, draw=black},
                every node near coord/.append style={
                    anchor=east,
                    rotate=90,
                    font=\footnotesize,
                }
            }}

    \begin{axis}[bar shift=-20pt]
        \addplot[fill=rwth-50] table [x expr=\coordindex, y=100us] {\resultsseq}; \uniquepgflabel{pgf:aoa_seq}
        \coordinate (topY) at (rel axis cs:0,1);
    \addplot[fill=rwth-25, bar width=2pt, postaction={solid, pattern={Lines[distance=1pt]}, pattern color=rwth-75}] table [x expr=\coordindex, y=100us] {\resultsdiff}; \uniquepgflabel{pgf:aoa_par}
    \end{axis}
    \begin{axis}[hide axis, bar shift=-16pt]
        \addplot[fill=grun-50] table [x expr=\coordindex, y=100us] {\refresultsseq}; \uniquepgflabel{pgf:avp_seq}
        \addplot[fill=grun-25, bar width=2pt, postaction={solid, pattern={Lines[distance=1pt]}, pattern color=grun}] table [x expr=\coordindex, y=100us] {\refresultsdiff}; \uniquepgflabel{pgf:avp_par}
    \end{axis}

    \begin{axis}[hide axis, bar shift=-11pt]
        \addplot[fill=rwth-50] table [x expr=\coordindex, y=500us] {\resultsseq};
        \addplot[fill=rwth-25, bar width=2pt, postaction={solid, pattern={Lines[distance=1pt]}, pattern color=rwth}] table [x expr=\coordindex, y=500us] {\resultsdiff};
    \end{axis}
    \begin{axis}[hide axis, bar shift=-7pt]
        \addplot[fill=grun-50] table [x expr=\coordindex, y=500us] {\refresultsseq};
        \addplot[fill=grun-25, bar width=2pt, postaction={solid, pattern={Lines[distance=1pt]}, pattern color=grun}] table [x expr=\coordindex, y=500us] {\refresultsdiff};
    \end{axis}

    \begin{axis}[hide axis, bar shift=-2pt]
        \addplot[fill=rwth-50] table [x expr=\coordindex, y=1ms] {\resultsseq};
        \addplot[fill=rwth-25, bar width=2pt, postaction={solid, pattern={Lines[distance=1pt]}, pattern color=rwth}] table [x expr=\coordindex, y=1ms] {\resultsdiff};
    \end{axis}
    \begin{axis}[hide axis, bar shift=+2pt]
        \addplot[fill=grun-50] table [x expr=\coordindex, y=1ms] {\refresultsseq};
        \addplot[fill=grun-25, bar width=2pt, postaction={solid, pattern={Lines[distance=1pt]}, pattern color=grun}] table [x expr=\coordindex, y=1ms] {\refresultsdiff};
    \end{axis}

    \begin{axis}[hide axis, bar shift=+7pt]
        \addplot[fill=rwth-50] table [x expr=\coordindex, y=5ms] {\resultsseq};
        \addplot[fill=rwth-25, bar width=2pt, postaction={solid, pattern={Lines[distance=1pt]}, pattern color=rwth}] table [x expr=\coordindex, y=5ms] {\resultsdiff};
    \end{axis}
    \begin{axis}[hide axis, bar shift=+11pt]
        \addplot[fill=grun-50] table [x expr=\coordindex, y=5ms] {\refresultsseq};
        \addplot[fill=grun-25, bar width=2pt, postaction={solid, pattern={Lines[distance=1pt]}, pattern color=grun}] table [x expr=\coordindex, y=5ms] {\refresultsdiff};
    \end{axis}

    \begin{axis}[hide axis, bar shift=+16pt]
        \addplot[fill=rwth-50] table [x expr=\coordindex, y=10ms] {\resultsseq};
        \addplot[fill=rwth-25, bar width=2pt, postaction={solid, pattern={Lines[distance=1pt]}, pattern color=rwth}] table [x expr=\coordindex, y=10ms] {\resultsdiff};
    \end{axis}
    \begin{axis}[hide axis, bar shift=+20pt]
        \addplot[fill=grun-50] table [x expr=\coordindex, y=10ms] {\refresultsseq};
        \addplot[fill=grun-25, bar width=2pt, postaction={solid, pattern={Lines[distance=1pt]}, pattern color=grun}] table [x expr=\coordindex, y=10ms] {\refresultsdiff};
    \end{axis}

    \begin{pgfonlayer}{background}
        \pgfplotstablegetrowsof{\resultsseq}
        \pgfmathsetmacro{\rcnt}{\pgfplotsretval-1}

        \begin{axis}[hide axis, ymin=0, ymax=1, xmin=0, xmax=\rcnt, stack plots=false]
            \foreach \X in {0,...,\rcnt} {
                \addplot[const plot, fill=rwth!01, update limits=false, draw=none] coordinates {(2*\X-.49, 1) (2*\X+0.49, 1)} \closedcycle;
                \addplot[const plot, fill=rwth!05, update limits=false, draw=none] coordinates {(2*\X-1-0.49, 1) (2*\X-1+0.49, 1)} \closedcycle;
            }
        \end{axis}
    \end{pgfonlayer}

    \coordinate (origin) at (0,0);
    \foreach \X in {0,...,3} {
        \node[font={\scriptsize}, rotate=90, anchor=east, inner sep=0pt] at ([xshift=-18pt, yshift=-4pt] xlabel\X.north|-origin) (sublabel_100us_\X) {\SI{0.1}{\milli\second}};
        \node[font={\scriptsize}, rotate=90, anchor=east, inner sep=0pt] at ([xshift=9pt] sublabel_100us_\X.east) (sublabel_500us_\X) {\SI{0.5}{\milli\second}};
        \node[font={\scriptsize}, rotate=90, anchor=east, inner sep=0pt] at ([xshift=9pt] sublabel_500us_\X.east) (sublabel_1ms_\X) {\SI{1}{\milli\second}};
        \node[font={\scriptsize}, rotate=90, anchor=east, inner sep=0pt] at ([xshift=9pt] sublabel_1ms_\X.east) (sublabel_5ms_\X) {\SI{5}{\milli\second}};
        \node[font={\scriptsize}, rotate=90, anchor=east, inner sep=0pt] at ([xshift=9pt] sublabel_5ms_\X.east) (sublabel_10ms_\X) {\SI{10}{\milli\second}};

        \draw[ultra thin, black] (origin-|sublabel_100us_\X) -- ++(0pt, -2pt);
        \draw[ultra thin, black] (origin-|sublabel_500us_\X) -- ++(0pt, -2pt);
        \draw[ultra thin, black] (origin-|sublabel_1ms_\X) -- ++(0pt, -2pt);
        \draw[ultra thin, black] (origin-|sublabel_5ms_\X) -- ++(0pt, -2pt);
        \draw[ultra thin, black] (origin-|sublabel_10ms_\X) -- ++(0pt, -2pt);
    }

    \matrix[
        draw,
        fill=white,
        matrix of nodes,
        font={\scriptsize},
        anchor=north west,
        row sep=1pt,
        column sep=2pt,
        inner sep=0.5pt,
        column 1/.style={anchor=base east},
        line width=0.3pt,
    ] at ([xshift=1pt,yshift=-1pt]topY) {
        VP\textbackslash Mode & Sequential & Parallel \\
        This Work & \uniquepgfref{pgf:aoa_seq} & \uniquepgfref{pgf:aoa_par} \\
        AVP64 & \uniquepgfref{pgf:avp_seq} & \uniquepgfref{pgf:avp_par} \\
    };
\end{tikzpicture}

    \caption{Bare-metal Dhrystone~\cite{weicker_dhrystone_1988} benchmark.}
    \label{fig:dhrystone}
\end{figure}
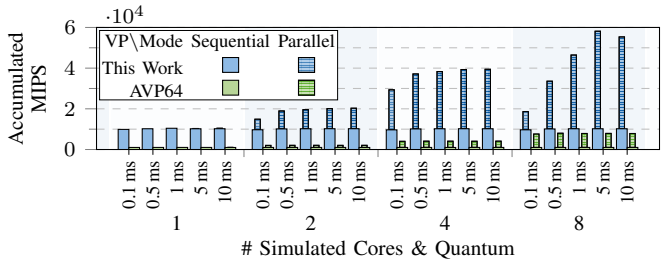

\Cref{fig:dhrystone} shows the measured accumulated \ac{mips} values for different core counts, quantum values, and parallelization settings.
For a single-core \ac{vp}, parallelization does not yield performance benefits as the \ac{cpu} is the only compute-intensive model that runs in the main thread.
Our \ac{aoa} \ac{vp} achieves nearly \SI{10000}{MIPS}, which is about \SI{10}{x} the performance of \ac{avp64}.

For dual-core setups with parallel execution enabled, the performance effectively doubles due to simultaneous simulation of both cores.
However, smaller quantum values lead to decreased \ac{aoa} performance due to increased synchronization overheads involving \ac{el}-switching for \ac{kvm} entries and exits~\cite{engblom_how_2023}.
This effect is even more visible in quad- and octa-core setups.

In octa-core configurations, limited host-machine performance cores (\num{6} in total) reduce the achievable speedups since some simulated cores have to run on efficiency cores.

In summary, the parallel Dhrystone benchmark shows that for compute-intensive workloads, the \ac{aoa} \ac{vp} achieves up to \SI{10}{x} speedup over the \ac{iss}-based \ac{avp64}.
Parallelized execution of the simulated cores significantly enhances the performance.
The optimal speedup can be reached for dual and quad-core setups.
For octa-core setups, the limited number of performance cores of the host machine reduces the achievable speedup.

\subsection{Linux's Boot Process}
\label{sec:results:boot}

The next benchmark we look at is the boot process of a Buildroot-based Linux~\cite{buildroot_making_2024}.
In contrast to the parallel Dhrystone, this benchmark represents a mostly sequential workload where one core performs the boot tasks while others idle.

\begin{figure}[!t]
    \centering

    \begin{subfigure}[b]{\linewidth}
        \centering
        \renewcommand{\uniquepgftag}{boot-normal}
        \pgfplotstableread[col sep=comma]{plot/data/aoavp_boot_seq.csv}\resultsseq
        \pgfplotstableread[col sep=comma]{plot/data/aoavp_boot_par.csv}\resultspar
        \pgfdeclarelayer{background}
\pgfsetlayers{background,main}

\begin{tikzpicture}
    \pgfplotsset{every axis/.style={ 
                BarConfig,
                bar width=4pt,
                height=3.2cm,
                ybar,
                ymin=0,
                minor y tick num=1,
                yminorgrids=true,
                major x tick style = {opacity=0},
                minor x tick num = 1,
                xticklabels from table={\resultsseq}{cores},
                x tick label style={yshift=-0.6cm, name=xlabel\ticknum},
                xticklabel style={rotate=0, anchor=north},
                ylabel={\shortstack{Wall-Clock\\Time (\si{\second})}},
                xlabel={\# Simulated Cores \& Quantum},
                xlabel shift=-3pt,
                enlarge x limits=0.2,
                legend style={at={([xshift=1pt,yshift=-1pt]0,1.0)}, anchor=north west, legend columns=4, inner sep=1pt, draw=black, font={\scriptsize}},
                legend image code/.code={\draw [#1] (0cm,-0.1cm) rectangle (0.2cm,0.1cm); },
                every node near coord/.append style={
                    anchor=east,
                    rotate=90,
                    font=\footnotesize,
                }
            }}
    \begin{axis}

        \addplot[fill=rwth-50, bar shift=-20pt] table [x expr=\coordindex, y=100us] {\resultsseq};
        \addlegendentry{Sequential}
        \addplot[fill=grun-50, bar shift=-16pt] table [x expr=\coordindex, y=100us] {\resultspar};
        \addlegendentry{Parallel}

        \addplot[fill=rwth-50, bar shift=-11pt] table [x expr=\coordindex, y=500us] {\resultsseq};
        \addplot[fill=grun-50, bar shift=-07pt] table [x expr=\coordindex, y=500us] {\resultspar};

        \addplot[fill=rwth-50, bar shift=-02pt] table [x expr=\coordindex, y=1ms] {\resultsseq};
        \addplot[fill=grun-50, bar shift=+02pt] table [x expr=\coordindex, y=1ms] {\resultspar};

        \addplot[fill=rwth-50, bar shift=+07pt] table [x expr=\coordindex, y=5ms] {\resultsseq};
        \addplot[fill=grun-50, bar shift=+11pt] table [x expr=\coordindex, y=5ms] {\resultspar};

        \addplot[fill=rwth-50, bar shift=+16pt] table [x expr=\coordindex, y=10ms] {\resultsseq};
        \addplot[fill=grun-50, bar shift=+20pt] table [x expr=\coordindex, y=10ms] {\resultspar};
    \end{axis}

    \begin{pgfonlayer}{background}
        \pgfplotstablegetrowsof{\resultsseq}
        \pgfmathsetmacro{\rcnt}{\pgfplotsretval-1}

        \begin{axis}[hide axis, ymin=0, ymax=1, xmin=0, xmax=\rcnt, stack plots=false]
            \foreach \X in {0,...,\rcnt} {
                \addplot[const plot, fill=rwth!01, update limits=false, draw=none] coordinates {(2*\X-.49, 1) (2*\X+0.49, 1)} \closedcycle;
                \addplot[const plot, fill=rwth!05, update limits=false, draw=none] coordinates {(2*\X-1-0.49, 1) (2*\X-1+0.49, 1)} \closedcycle;
            }
        \end{axis}
    \end{pgfonlayer}

    \coordinate (origin) at (0,0);
    \foreach \X in {0,...,3} {
        \node[font={\scriptsize}, rotate=90, anchor=east, inner sep=0pt] at ([xshift=-18pt, yshift=-4pt] xlabel\X.north|-origin) (sublabel_100us_\X) {\SI{0.1}{\milli\second}};
        \node[font={\scriptsize}, rotate=90, anchor=east, inner sep=0pt] at ([xshift=9pt] sublabel_100us_\X.east) (sublabel_500us_\X) {\SI{0.5}{\milli\second}};
        \node[font={\scriptsize}, rotate=90, anchor=east, inner sep=0pt] at ([xshift=9pt] sublabel_500us_\X.east) (sublabel_1ms_\X) {\SI{1}{\milli\second}};
        \node[font={\scriptsize}, rotate=90, anchor=east, inner sep=0pt] at ([xshift=9pt] sublabel_1ms_\X.east) (sublabel_5ms_\X) {\SI{5}{\milli\second}};
        \node[font={\scriptsize}, rotate=90, anchor=east, inner sep=0pt] at ([xshift=9pt] sublabel_5ms_\X.east) (sublabel_10ms_\X) {\SI{10}{\milli\second}};

        \draw[ultra thin, black] (origin-|sublabel_100us_\X) -- ++(0pt, -2pt);
        \draw[ultra thin, black] (origin-|sublabel_500us_\X) -- ++(0pt, -2pt);
        \draw[ultra thin, black] (origin-|sublabel_1ms_\X) -- ++(0pt, -2pt);
        \draw[ultra thin, black] (origin-|sublabel_5ms_\X) -- ++(0pt, -2pt);
        \draw[ultra thin, black] (origin-|sublabel_10ms_\X) -- ++(0pt, -2pt);
    }
\end{tikzpicture}
        \vspace{-1em}
        \caption{Without \acs{wfi} annotation.}
        \label{fig:boot:wowfi}
    \end{subfigure}

    \begin{subfigure}[b]{\linewidth}
        \centering
        \renewcommand{\uniquepgftag}{boot-wfi}
        \pgfplotstableread[col sep=comma]{plot/data/aoavp_boot_seq_wfi.csv}\resultsseq
        \pgfplotstableread[col sep=comma]{plot/data/aoavp_boot_par_wfi.csv}\resultspar
        \pgfdeclarelayer{background}
\pgfsetlayers{background,main}

\begin{tikzpicture}
    \pgfplotsset{every axis/.style={ 
                BarConfig,
                bar width=4pt,
                height=3.2cm,
                ybar,
                ymin=0,
                minor y tick num=1,
                yminorgrids=true,
                major x tick style = {opacity=0},
                minor x tick num = 1,
                xticklabels from table={\resultsseq}{cores},
                x tick label style={yshift=-0.6cm, name=xlabel\ticknum},
                xticklabel style={rotate=0, anchor=north},
                ylabel={\shortstack{Wall-Clock\\Time (\si{\second})}},
                xlabel={\# Simulated Cores \& Quantum},
                xlabel shift=-3pt,
                enlarge x limits=0.2,
                legend style={at={([xshift=1pt,yshift=-1pt]0,1.0)}, anchor=north west, legend columns=4, inner sep=1pt, draw=black, font={\scriptsize}},
                legend image code/.code={\draw [#1] (0cm,-0.1cm) rectangle (0.2cm,0.1cm); },
                every node near coord/.append style={
                    anchor=east,
                    rotate=90,
                    font=\footnotesize,
                }
            }}
    \begin{axis}

        \addplot[fill=rwth-50, bar shift=-20pt] table [x expr=\coordindex, y=100us] {\resultsseq};
        \addlegendentry{Sequential}
        \addplot[fill=grun-50, bar shift=-16pt] table [x expr=\coordindex, y=100us] {\resultspar};
        \addlegendentry{Parallel}

        \addplot[fill=rwth-50, bar shift=-11pt] table [x expr=\coordindex, y=500us] {\resultsseq};
        \addplot[fill=grun-50, bar shift=-07pt] table [x expr=\coordindex, y=500us] {\resultspar};

        \addplot[fill=rwth-50, bar shift=-02pt] table [x expr=\coordindex, y=1ms] {\resultsseq};
        \addplot[fill=grun-50, bar shift=+02pt] table [x expr=\coordindex, y=1ms] {\resultspar};

        \addplot[fill=rwth-50, bar shift=+07pt] table [x expr=\coordindex, y=5ms] {\resultsseq};
        \addplot[fill=grun-50, bar shift=+11pt] table [x expr=\coordindex, y=5ms] {\resultspar};

        \addplot[fill=rwth-50, bar shift=+16pt] table [x expr=\coordindex, y=10ms] {\resultsseq};
        \addplot[fill=grun-50, bar shift=+20pt] table [x expr=\coordindex, y=10ms] {\resultspar};
    \end{axis}

    \begin{pgfonlayer}{background}
        \pgfplotstablegetrowsof{\resultsseq}
        \pgfmathsetmacro{\rcnt}{\pgfplotsretval-1}

        \begin{axis}[hide axis, ymin=0, ymax=1, xmin=0, xmax=\rcnt, stack plots=false]
            \foreach \X in {0,...,\rcnt} {
                \addplot[const plot, fill=rwth!01, update limits=false, draw=none] coordinates {(2*\X-.49, 1) (2*\X+0.49, 1)} \closedcycle;
                \addplot[const plot, fill=rwth!05, update limits=false, draw=none] coordinates {(2*\X-1-0.49, 1) (2*\X-1+0.49, 1)} \closedcycle;
            }
        \end{axis}
    \end{pgfonlayer}

    \coordinate (origin) at (0,0);
    \foreach \X in {0,...,3} {
        \node[font={\scriptsize}, rotate=90, anchor=east, inner sep=0pt] at ([xshift=-18pt, yshift=-4pt] xlabel\X.north|-origin) (sublabel_100us_\X) {\SI{0.1}{\milli\second}};
        \node[font={\scriptsize}, rotate=90, anchor=east, inner sep=0pt] at ([xshift=9pt] sublabel_100us_\X.east) (sublabel_500us_\X) {\SI{0.5}{\milli\second}};
        \node[font={\scriptsize}, rotate=90, anchor=east, inner sep=0pt] at ([xshift=9pt] sublabel_500us_\X.east) (sublabel_1ms_\X) {\SI{1}{\milli\second}};
        \node[font={\scriptsize}, rotate=90, anchor=east, inner sep=0pt] at ([xshift=9pt] sublabel_1ms_\X.east) (sublabel_5ms_\X) {\SI{5}{\milli\second}};
        \node[font={\scriptsize}, rotate=90, anchor=east, inner sep=0pt] at ([xshift=9pt] sublabel_5ms_\X.east) (sublabel_10ms_\X) {\SI{10}{\milli\second}};

        \draw[ultra thin, black] (origin-|sublabel_100us_\X) -- ++(0pt, -2pt);
        \draw[ultra thin, black] (origin-|sublabel_500us_\X) -- ++(0pt, -2pt);
        \draw[ultra thin, black] (origin-|sublabel_1ms_\X) -- ++(0pt, -2pt);
        \draw[ultra thin, black] (origin-|sublabel_5ms_\X) -- ++(0pt, -2pt);
        \draw[ultra thin, black] (origin-|sublabel_10ms_\X) -- ++(0pt, -2pt);
    }
\end{tikzpicture}

        \caption{With \acs{wfi} annotations.}
        \label{fig:boot:wfi}
    \end{subfigure}

    \caption{Buildroot Linux boot durations for \acs{aoa}.}
    \label{fig:boot}
\end{figure}
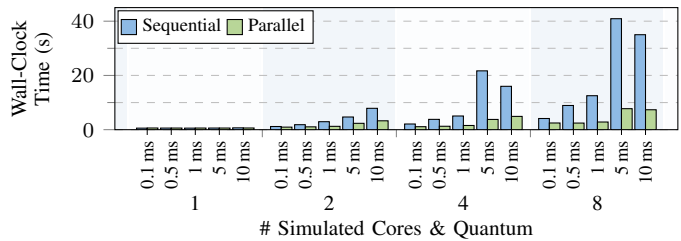
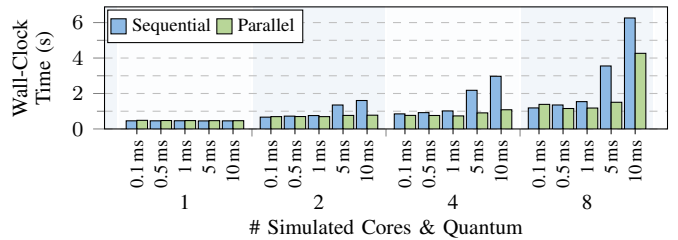

\Cref{fig:boot:wowfi} illustrates the wall-clock time required for different core counts without \ac{wfi} annotations (see \cref{sec:implementation:wfi}, \ac{wfi} is handled by \ac{kvm}).
While the single-core \ac{vp} boots Linux in approximately \SI{0.6}{\second}, multicore setups take considerably longer.
A non-parallelized boot of an octa-core \ac{vp} can take up to \SI{40}{\second}.
The reason for this is that in addition to the core that performs the boot, the idle loops of the other cores need to be simulated.
For larger quantum values, synchronization between the cores is more complicated which leads to increased runtime.
This has also been observed by previous works~\cite{zurstrasen_optimal_2024,zurstrasen_art_2025,engblom_notes_2022}.

When parallelization is activated, the idling cores can be simulated in parallel which reduces the needed amount of wall-clock time.
However, booting a multicore system is still slower than booting a single-core one due to the additional overhead.

Apart from applying parallelization, another way to optimize the idle-loop behavior is to annotate the \ac{wfi} instruction (see \cref{sec:implementation:wfi}).
\Cref{fig:boot:wfi} shows how \ac{wfi} annotations effectively increase the performance of the Linux boot.
Instead of simulating the idle loop, the simulation of the core is suspended until an interrupt is signaled.
This reduces the time needed for a Linux boot to under \SI{1}{\second} for dual and quad-core setups.
For the octa-core \ac{vp}, speedups achieved by \ac{wfi} annotation range from \SI{1.78}{x} for the \SI{100}{\micro\second} parallel version up to \SI{11.5}{x} for the \SI{5}{\milli\second} sequential version.

The Linux-boot benchmark shows that the simulation of idle loops drastically reduces the performance of the sequential simulation.
Parallelizing the simulation helps to counteract this effect.
However, the best results are achieved when idle loops are not simulated but annotated, so the simulation of the core can be suspended.
When both techniques are combined, the needed time for a multicore Linux boot can be kept small.

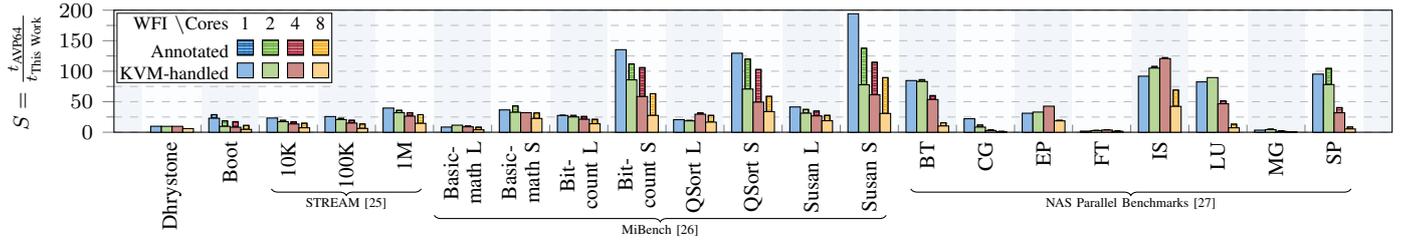
\begin{figure*}[!t]
    \centering

    \renewcommand{\uniquepgftag}{bench-1ms}
    \pgfplotstableread[col sep=comma]{plot/data/aoavp_1ms_no_wfi_speedup.csv}\result
    \pgfplotstableread[col sep=comma]{plot/data/aoavp_1ms_diff_speedup.csv}\resultdiffwfi
    \pgfdeclarelayer{background}
\pgfsetlayers{background,main}

\begin{tikzpicture}
    \pgfplotsset{every axis/.style={ 
                BarConfig,
                bar width=4pt,
                ybar stacked,
                ymin=0,
                ymax=200,
                yminorgrids=true,
                minor x tick num = 1,
                minor y tick num = 1,
                xtick align=inside,
                major x tick style = {opacity=0},
                xticklabels from table={\result}{benchmark},
                x tick label style={name=xlabel\ticknum},
                xticklabel style={rotate=90, anchor=east},
                ylabel={$S=\frac{t_{\text{AVP64}}}{t_{\text{This Work}}}$},
                enlarge x limits=0.05,
                legend style={at={([yshift=-1pt]0.5,1.0)}, anchor=north, legend columns=4, inner sep=1pt, draw=black},
            }}

    \begin{axis}[bar shift=-6pt]
        \addplot[fill=rwth-50] table [x expr=\coordindex, y=1] {\result}; \uniquepgflabel{pgf:1}
        \addplot[fill=rwth-50, bar width=2pt, postaction={solid, pattern={Lines[distance=1pt]}, pattern color=rwth}] table [x expr=\coordindex, y=1] {\resultdiffwfi}; \uniquepgflabel{pgf:1wfi}
        \coordinate (topY) at (rel axis cs:0,1);
    \end{axis}

    \begin{axis}[bar shift=-2pt, hide axis]
        \addplot[fill=grun-50] table [x expr=\coordindex, y=2] {\result}; \uniquepgflabel{pgf:2}
        \addplot[fill=grun-50, bar width=2pt, postaction={solid, pattern={Lines[distance=1pt]}, pattern color=grun}] table [x expr=\coordindex, y=2] {\resultdiffwfi}; \uniquepgflabel{pgf:2wfi}
    \end{axis}

    \begin{axis}[bar shift=+2pt, hide axis]
        \addplot[fill=bordeaux-50] table [x expr=\coordindex, y=4] {\result}; \uniquepgflabel{pgf:4}
        \addplot[fill=bordeaux-50, bar width=2pt,  postaction={solid, pattern={Lines[distance=1pt]}, pattern color=bordeaux}] table [x expr=\coordindex, y=4] {\resultdiffwfi}; \uniquepgflabel{pgf:4wfi}
    \end{axis}

    \begin{axis}[bar shift=+6pt, hide axis]
        \addplot[fill=orange-50] table [x expr=\coordindex, y=8] {\result}; \uniquepgflabel{pgf:8}
        \addplot[fill=orange-50, bar width=2pt, postaction={solid, pattern={Lines[distance=1pt]}, pattern color=orange}] table [x expr=\coordindex, y=8] {\resultdiffwfi}; \uniquepgflabel{pgf:8wfi}
    \end{axis}


    \begin{pgfonlayer}{background}
        \pgfplotstablegetrowsof{\result}
        \pgfmathsetmacro{\rcnt}{\pgfplotsretval-1}

        \begin{axis}[hide axis, ymin=0, ymax=1, xmin=0, xmax=\rcnt, stack plots=false]
            \foreach \X in {0,...,11} {
                \addplot[const plot, fill=rwth!01, update limits=false, draw=none] coordinates {(2*\X-.49, 1) (2*\X+0.49, 1)} \closedcycle;
                \addplot[const plot, fill=rwth!05, update limits=false, draw=none] coordinates {(2*\X-1-0.49, 1) (2*\X-1+0.49, 1)} \closedcycle;
            }
        \end{axis}
    \end{pgfonlayer}

    \path[draw, decorate, decoration=brace] ([yshift=-0.75cm]xlabel4.south east) -- ([yshift=-0.75cm]xlabel2.north east) node[midway,below,font=\tiny]{STREAM~\cite{mccalpin_memory_1995}};
    \path[draw, decorate, decoration=brace] ([yshift=-1.1cm]xlabel12.south east) -- ([yshift=-1.1cm]xlabel5.north east) node[midway,below,font=\tiny]{MiBench~\cite{guthaus_mibench_2001}};
    \path[draw, decorate, decoration=brace] ([yshift=-0.75cm]xlabel20.south east) -- ([yshift=-0.75cm]xlabel13.north east) node[midway,below,font=\tiny]{\acl{npb}~\cite{nasa_jet_propulsion_laboratory_nasa-jplembedded-gcov_2023}};

    \matrix[
        draw,
        fill=white,
        matrix of nodes,
        font={\scriptsize},
        anchor=north west,
        row sep=1pt,
        column sep=2pt,
        inner sep=0.5pt,
        column 1/.style={anchor=base east},
        line width=0.3pt,
    ] at ([xshift=1pt, yshift=-1pt]topY) {
        WFI \textbackslash Cores & 1 & 2 & 4 & 8 \\
        Annotated & \uniquepgfref{pgf:1wfi} & \uniquepgfref{pgf:2wfi} & \uniquepgfref{pgf:4wfi} & \uniquepgfref{pgf:8wfi} \\
        \acs{kvm}-handled & \uniquepgfref{pgf:1} & \uniquepgfref{pgf:2} & \uniquepgfref{pgf:4} & \uniquepgfref{pgf:8} \\
    };

\end{tikzpicture}
    \vspace{-1em}

    \caption{Benchmark speedup $S$ of \ac{aoa} compared to \ac{avp64} for \SI{1}{\milli\second} quantum and activated parallel execution.}
    \label{fig:bench}
    \vspace{-2em}
\end{figure*}

\subsection{User-Space Benchmarks}

After analyzing fully parallelizable and a predominantly sequential workloads, we now look at common user-space benchmarks executed within a Linux environment.
The results are depicted in \cref{fig:bench} for a \SI{1}{\milli\second} quantum with parallel execution enabled, and various core configurations.
All listed benchmarks are executed on \ac{avp64} and our \ac{aoa} \ac{vp}.
The wall-clock time it takes to execute the benchmarks is measured.
On the y-axis of \cref{fig:bench}, one can see the speedup of \ac{aoa} compared to the execution on a similarly configured \ac{avp64} (quantum, number of cores, parallelized execution).

In addition to the user space benchmarks, we also include results for the bare-metal Dhrystone and Linux-boot process.
As observed in \cref{sec:results:dhrystone}, the Dhrystone benchmark shows a dip in speedup for eight simulated cores due to the number of available performance cores.
For the Linux boot process, increased core counts reduce the speedup because trapping \ac{wfi} instructions is less expensive for \acp{iss} than for \ac{aoa} due to the needed context switching for exiting \ac{kvm}~\cite{engblom_how_2023}.

\subsubsection{STREAM}
The first user space benchmark tested is \textit{STREAM}~\cite{mccalpin_memory_1995}, which measures memory bandwidth by executing numerous load and store instructions~\cite{bosbach_entropy-based_2023}.
We run the benchmark with different array sizes ranging from \num{10000} (\textit{10K}) to \num{1000000} (\textit{1M}) elements.
For \ac{iss}-based simulators like \ac{avp64}, these instructions require \ac{sw}-based \ac{mmu} translations from virtual to physical addresses.
This task incurs significant overhead.
In contrast, \ac{aoa} models leverage the \ac{hw}-provided two-stage address translation process of the host \ac{mmu} that handles these translations natively~\cite{arm_arm_2024,arm_ltd_armv8-virtualization_2019}.

\subsubsection{MiBench Suite}
Next, we evaluate benchmarks from the \textit{MiBench}~\cite{guthaus_mibench_2001} suite, representing automotive and industrial control workloads.
Theses single-threaded benchmarks do not benefit from a multicore setup.
Therefore, annotating \ac{wfi} instructions in the idle loop is essential to limit the performance loss on multicore setups.
Speedups compared to the \ac{iss}-based \ac{avp64} range significantly from approximately \SI{8}{x} for \textit{Basicmath L} up to \num{165}{x} for \textit{Susan S} on single-core \acp{vp}.

MiBench provides both large (\textit{L}) and small (\textit{S}) versions of each benchmark type.
The variants only differ in size of the input but not in the task itself.
It can be observed that the smaller variants achieve higher speedups than the larger ones.
This indicates that the huge speedup obtained by the small variants is not mainly caused by the computation itself.
For the \ac{dbt}-based \ac{iss} used by \ac{avp64}, basic blocks of the target \ac{sw} are translated to the host \ac{isa}.
Once they have been translated, they are cached, so the translation is only needed once.
When the size of the input for a workload is increased, usually the loops of the algorithm are just executed more often.
That means the translation overhead of the \ac{iss} has a larger impact when the translated blocks are executed less often.
The \ac{iss} therefore gets more efficient for increased workload-input sizes.
For \ac{aoa}, those effects do not exist whereby the speedup of \ac{aoa} compared to \ac{avp64} is reduced when the \ac{iss} gets more efficient.

\subsubsection{\acl{npb}}
Lastly, we examine multicore benchmarks from the \ac{npb}~\cite{nasa_jet_propulsion_laboratory_nasa-jplembedded-gcov_2023} suite.
These benchmarks distribute their workloads across multiple threads using \ac{openmp}~\cite{chandra_parallel_2001}.
Since the benchmarks use all cores of the simulated system, the system does not spend much time in idle mode.
Therefore, \ac{wfi} annotation does not significantly improve the performance.

Workloads that involve a lot of synchronization and therefore communication between the cores, like the \textit{CG}, \textit{FT}, and \textit{MG} benchmarks, cause more overhead than the other workloads.
However, with a minimum speedup of \SI{1.8}{x} for the FT benchmark, \ac{aoa} is still faster than \ac{avp64}.

In summary, the benchmark results show that \ac{aoa} is a promising approach that results in huge speedups compared to traditional \ac{iss}-based \acp{vp}.
The achieved speedup of \SI{2.57}{x} of \cite{junger_arm--arm_2020} for a bare-metal Coremark compared to an \ac{iss}-based \ac{vp} can be increased to up to \SI{10}{x} for Dhrystone on modern ARM \ac{hw} with parallel execution.
Apart from the observed speedups, the results show that idle-loop annotations or \ac{wfi} trapping are essential to limit the performance loss of single-threaded workloads on a multicore \ac{vp}.

\section{Conclusion \& Future Work}

In this paper, we presented a novel approach to \ac{aoa} virtualization within SystemC-based \acp{vp}, leveraging Linux's \ac{kvm} to enhance performance.
By running the target \ac{sw} natively on ARM-based hosts with \ac{vhe}, we eliminate the need for an \acp{iss} and thereby significantly improve the simulation performance.
Our SystemC-\ac{tlm}-based multicore \ac{cpu} model operates independently of host performance counters or custom kernel patches, which makes it more independent of the environment.

Benchmark results demonstrated that our \ac{aoa}-based \ac{vp} approach achieves substantial speedups compared to traditional \ac{iss}-based models.
For compute-intensive workloads, such as parallel Dhrystone, our solution reaches up to \SI{10}{x} the performance over an \ac{iss}-based \ac{vp}.
Additionally, by annotating \ac{wfi} instructions, we were able to further optimize the simulation of idle loops during processes like Linux booting, achieving significant reductions in required wall-clock time.

Overall, our work shows that \ac{aoa} virtualization is a promising technique for \ac{sw} development and testing.
We presented that the annotation of idle loops improves the simulation performance of certain workloads.
While our annotation-based approach works well for Linux-based target \ac{sw}, it needs to be adapted for other workloads.
For future work, we plan to further improve or automate this way of trapping \ac{wfi} instructions to be able to handle \ac{wfi} instructions without manual annotations.

Another limitation of our approach is that \ac{kvm} currently only supports \ac{el}0 and \ac{el}1 for the guest.
This is sufficient for most workloads including \acp{os} and user-space \ac{sw}.
However, hypervisors running in \ac{el}2 and ARM's trusted firmware running in \ac{el}3 cannot be executed with our approach.

In addition, instruction emulation can be added to support new instructions that may emerge in the future that are not supported by the host.
Furthermore, the approach can be extended to other architectures that have a virtualization extension, such as RISC-V-on-RISC-V simulation.

\newpage

\balance
\bibliographystyle{IEEEtran}
\bibliography{IEEEabrv,bstcontrol.bib,library.bib}

\end{document}